\documentclass[10pt,twocolumn,letterpaper]{article}

\usepackage{cvpr}              % To produce the CAMERA-READY version
\usepackage[dvipsnames]{xcolor}

\definecolor{cvprblue}{rgb}{0.21,0.49,0.74}
\usepackage[pagebackref,breaklinks,colorlinks,citecolor=cvprblue]{hyperref}
\usepackage{overpic}
\usepackage{pifont}
\usepackage{graphicx} 
\usepackage{multirow}
\usepackage{makecell}
\usepackage{wasysym}
\usepackage[accsupp]{axessibility} % Improves PDF readability for those with visual impairments.
\def\paperID{6320} % *** Enter the Paper ID here
\def\confName{CVPR}
\def\confYear{2024}

\newcommand{\authorskip}{\hspace{5mm}}

\title{A Stealthy Wrongdoer: Feature-Oriented Reconstruction Attack \\ against Split Learning}

\author{Xiaoyang Xu$^{1}$ \authorskip Mengda Yang$^{1}$ \authorskip Wenzhe Yi $^{1}$ \authorskip Ziang Li $^{1}$ 
\authorskip Juan Wang$^{1}$\thanks{Corresponding author.} \authorskip Hongxin Hu $^{2}$ \\ Yong Zhuang $^{1}$ \authorskip Yaxin Liu $^{1}$\\
$^{1}$ Key Laboratory of Aerospace Information Security and Trusted Computing, Ministry of Education, \\ School of Cyber Science and Engineering, Wuhan University\\
$^{2}$ Department of Computer Science and Engineering, University at Buffalo, SUNY
\\
{\tt\small {\{xiaoyangx, mengday, wenzhey, ziangli, yong.zhuang, yaxin.liu\}@whu.edu.cn}}
\\ 
{\tt\small jwang@whu.edu.cn,  hongxinh@buffalo.edu}
}
\begin{document}
\maketitle
\begin{abstract}
Split Learning (SL) is a distributed learning framework renowned for its privacy-preserving features and minimal computational requirements. Previous research consistently highlights the potential privacy breaches in SL systems by server adversaries reconstructing training data. However, these studies often rely on strong assumptions or compromise system utility to enhance attack performance. This paper introduces a new semi-honest Data Reconstruction Attack on SL, named Feature-Oriented Reconstruction Attack (FORA). In contrast to prior works, FORA relies on limited prior knowledge, specifically that the server utilizes auxiliary samples from the public without knowing any client's private information. This allows FORA to conduct the attack stealthily and achieve robust performance. The key vulnerability exploited by FORA is the revelation of the model representation preference in the smashed data output by victim client. FORA constructs a substitute client through feature-level transfer learning, aiming to closely mimic the victim client's representation preference. Leveraging this substitute client, the server trains the attack model to effectively reconstruct private data. Extensive experiments showcase FORA's superior performance compared to state-of-the-art methods. Furthermore, the paper systematically evaluates the proposed method's applicability across diverse settings and advanced defense strategies.
\renewcommand{\thefootnote}{\fnsymbol{footnote}}
\setcounter{footnote}{-1}
% Our code is available at https://github.com/X1aoyangXu/FORA.
\footnote{Our code is available at https://github.com/X1aoyangXu/FORA}

\end{abstract}
    
\section{Introduction}
\label{sec:Introduction}
Deep Neural Networks (DNN) have gained widespread usage in computer vision due to their excellent learning ability and expressive power. Split Learning (SL) \cite{abuadbba2020can,gupta2018distributed,poirot2019split,vepakomma2018split,vepakomma2020nopeek,gao2020end,thapa2022splitfed} emerged as a distributed collaborative framework that enables clients to cooperate with a server to perform learning task. 
In SL, the complete DNN model is divided into two parts, which are deployed on the client and server respectively. For a normal training process in SL, the client performs the computational process locally and communicates with the server solely based on intermediate features (referred to as smashed data) and their corresponding gradients. In this case, the server does not have access to any private information (raw data, parameters, architecture) about the client. Therefore, SL is considered effective in protecting the privacy of clients.

However, recent works \cite{he2019model, gaopcat,pasquini2021unleashing,erdougan2022unsplit,song2019overlearning} have shown that there are still privacy risks associated with SL. It is possible for the server to steal private information about the client according to auxiliary knowledge. One particular concern is the Data Reconstruction Attack (DRA) \cite{gaopcat,pasquini2021unleashing,erdougan2022unsplit}, where a server attempts to recover the training data of a client in SL systems.
Depending on whether the server affects the normal process of SL, we can categorize adversaries into malicious and semi-honest attackers. 
Malicious servers such as FSHA \cite{pasquini2021unleashing} can manipulate the SL training process to conduct more effective attack. However, the latest findings ~\cite{hust_ndss,Splitguard} show that FSHA's mischief is easily detected by the client, leading to the termination of SL training protocol
For semi-honest attackers, \eg PCAT \cite{gaopcat} and UnSplit \cite{erdougan2022unsplit}, their superior camouflage makes them less likely to be detected. But current semi-honest attackers often rely overly on assumptions that favor their performances. For example, UnSplit requires knowledge of the client's architecture and is only applicable to simple networks or datasets.
As for PCAT,  it unduly depends on the availability of partial private data to assist in training the pseudo-client. 
These assumptions contradict the basic principle of SL, which is to ensure that the client's knowledge remains hidden from the server.
% In summary, when we review previous attacks, we find that they lack consideration of the intrinsic security of SL and the plausibility of their attack hypothesis, which limits the effectiveness and threat of their approach in real-world SL systems scenarios.
In summary, we find previous attacks lack consideration of the intrinsic security of SL and the plausibility of their attack hypothesis, which limits the effectiveness and threat of their approach in real-world SL systems scenarios.

In this work, we introduce a novel DRA toward more realistic and more challenging scenarios, where the server cannot access private data or  structures and parameters of the client model. Our scheme stems from new insights into potential privacy breaches in SL. We discover a fundamental phenomenon that the client model has its own \emph{representation preference}, which can be reflected through the output smashed data. More importantly, this unique information can indicate the feature extraction behavior of the client.
Based on this new insight, we propose a semi-honest privacy threat, namely Feature-Oriented Reconstruction Attack (FORA). A server adversary could establish a substitute client by narrowing the reference distance with the real client, which allows the substitute model to mimic the behavior of the target model at a finer granularity. To efficiently measure the preference distance of different representations, we introduce domain Discriminator network \cite{goodfellow2014generative,discriminator_da} and Multi-Kernel Maximum Mean Discrepancy (MK-MMD) \cite{mkmmd1,mkmmd2}. These techniques are widely used in domain adaptation \cite{ad_survey}, enabling us to project various representation preferences into a shared space for comparison. With a well-trained substitute client, the server can successfully recover the private data by constructing an inverse network.
% It is worth noting that we do not rely on any unreasonable adversary assumptions, such as access to private training data or prior knowledge of the target client model. All we need is an auxiliary dataset that is from the same domain  but with a different distribution. Such data can simply be collected from public sources.

We conduct our evaluation on two benchmark datasets and corresponding networks against different model partitioning strategies. The experimental results indicate that the proposed method significantly outperforms baseline attacks.
% Taking the reconstructed images of CelebA at layer 2 as an example, FORA achieves 2.8x and 1.5x improvement on  SSIM \cite{ssim}, 1.97x and 1.42x improvement on PSNR \cite{psnr}, 1.96x and 1.71x improvement on LPIPS \cite{lpips} respectively compared to UnSplit and PCAT. 
Taking the reconstructed images of CelebA at layer 2 as an example, UnSplit, PCAT and FORA achieve effects of 8.70, 12.05, and 17.11 on the PSNR \cite{psnr}.
This demonstrates that FORA has significantly outperformed  by 1.97x and 1.42x compared to the other two attacks.
% FORA achieves a notable improvement of 1.97x and 1.42x, respectively, compared to other attacks.
Although FSHA can achieve attack performance similar to ours, its malicious attack process can be promptly halted through monitoring mechanisms \cite{hust_ndss}, resulting in poor reconstructions. Furthermore, we investigate the potential influences on FORA, including different public knowledge conditions and existing defense strategies, to validate the robustness of FORA.
% Our additional ablation experiments reveal that even in scenarios where the distribution of the auxiliary dataset significantly differs from that of the private set, FORA can satisfactorily reconstruct the training data by solely relying on four categories of auxiliary data, with only a 0.062 drop in SSIM.

The main contributions of this paper can be summarized as follows:
% Our code is available at https://github.com/X1aoyangXu/FORA.
\begin{itemize}
\item[$\bullet$]
We propose a novel attack, named Feature-Oriented Reconstruction Attack (FORA). As far as we know, FORA is the first work enabling a semi-honest server to perform powerful DRA in more realistic and challenging SL systems. In such scenarios, the server has no prior knowledge of the client model or access to raw data.
\item[$\bullet$]
We have uncovered an inherent vulnerability in SL, where the server can exploit rich information in the smashed data to steal client representation preference, thereby building a substitute client for better reconstruction.
% We have uncovered inherent security vulnerabilities in SL, where the smashed data from a client model can reveal its representation preference. Based on this insight, a server adversary can start from the feature level, and train a substitute client in a fine-grained manner to perform data reconstruction attack. 
\item[$\bullet$]
We conduct comprehensive experiments with various adversarial knowledge against different benchmark datasets and models. The results demonstrate that FORA can achieve state-of-the-art attack performance compared with baselines and exhibits notable robustness across different settings.
\end{itemize}

\section{Background and Related Work}
\label{sec:Background and Related Work}

\begin{figure}[t]
\centering
\includegraphics[width=3in]{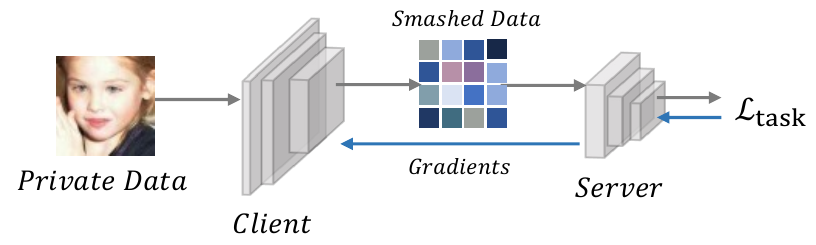}
\caption{Architecture of two-part split learning.}
\label{fig:sl}
\end{figure}

\textbf{Split Learning (SL).}
SL \cite{abuadbba2020can,vepakomma2018split,gupta2018distributed,poirot2019split,thapa2022splitfed} is an emerging distributed learning paradigm for resource-limited scenarios, which can split the neural network model into both client-side and server-side. As shown in \cref{fig:sl}, the client performs forward propagation and transmits the smashed data to the server, which then uses the computed loss for backward propagation and sends the gradients of the smashed data back to the client. Both the client and server will update their weights after receiving the gradients. It is generally believed that SL provides a secure and efficient training protocol by allowing the client to retain a portion of the model and training data locally while offloading most of the computing overhead to the server \cite{abuadbba2020can,gupta2018distributed,poirot2019split,vepakomma2018split}. However, recent studies \cite{pasquini2021unleashing,gaopcat,erdougan2022unsplit,fu2022label,kariyappa2023exploit} have highlighted vulnerabilities in SL, where the server can exploit the latter part of the model to carry out privacy attacks.

\textbf{Data Reconstruction Attack (DRA) on SL.}
DRA \cite{sfd,he2019model,disco, glass} is one of the most powerful privacy attacks that aim to steal the input data by the model's intermediate features. In SL, the server can utilize the smashed data output by the client to reconstruct the training data \cite{erdougan2022unsplit,gaopcat,pasquini2021unleashing}. One notable attack is known as FSHA \cite{pasquini2021unleashing}, where a malicious attacker utilizes the elaborated loss to alter the feature space of the victim client for reconstructing private data. In UnSplit \cite{erdougan2022unsplit}, the semi-honest server attempts to reconstruct the training data and client's parameters simultaneously by utilizing the smashed data. Specifically, UnSplit optimizes parameters and inputs sequentially by minimizing the outputs between the clone client and the target client. To the best of our knowledge, PCAT \cite{gaopcat} represents the most advanced attack under the semi-honest assumption. PCAT leverages the knowledge embedded in various stages of the server models to steal private data by constructing a pseudo-client.
Unlike previous work, SFA \cite{luo2023feature} focuses on reconstructing samples during the inference stage rather than the training samples.  

% \begin{table}[tbp]
% \centering
% \caption{Threat models among different attacks.}
% \resizebox{\linewidth}{!}{
% \begin{tabular}{c |c c c c}
%     \toprule
%     \textbf{Attacks} & \textbf{\makecell{Utility \\Disruption}} & \textbf{\makecell{Knowledge of \\Model Structure}} & \textbf{\makecell{Access to \\Private Data}} & \textbf{\makecell{Knowledge of \\Learning Task}}\\
%     \midrule
%     FSHA \cite{pasquini2021unleashing} & $\checkmark$ & $\times$ & $\times$ & $\times$\\
%     UnSplit \cite{erdougan2022unsplit}& $\times$ & $\checkmark$ & $\times$ & $\checkmark$\\
%     PCAT \cite{gaopcat}& $\times$ & $\times$ & $\checkmark$ & $\checkmark$\\
%     \textbf{FORA} & $\times$ & $\times$ & $\times$ & $\times$\\
%     \bottomrule
% \end{tabular}
% }
% \label{tab:comparison}
% \end{table}

Although existing works claim that their attacks pose significant privacy threats to SL, they disregard the plausibility of their threat model. For FSHA, the server reconstructs the raw data while at the cost of destroying the client's utility. While FSHA assumes that the client is entirely free of any awareness of being maliciously disrupted, recent research \cite{Splitguard, hust_ndss} indicates that such a malicious server can be easily detected by the client, leading to a halt in the SL. UnSplit needs the knowledge of the client's structure and is not suitable for complex networks and datasets due to the infinite searching space of input data and model parameters. As for PCAT, it requires the adversary to have access to a portion of the private dataset. This is an unreasonable assumption that violates the original intention of SL since one of the distinctive characteristics of SL is the ability to train models without sharing the raw data \cite{vepakomma2018split}. As a result, how to explore DRA under more realistic assumptions in SL remains an open question.
% In \cref{tab:comparison}, we show the threat models among different attacks.

\textbf{Domain Adaptation.}
Domain adaptation \cite{mmd1,mmd2,mkmmd1,mkmmd2,discriminator_da,wang2018deep,discriminator_da2} is a technique that seeks to enhance the generalization of a model by transferring knowledge acquired from a source domain to a distinct yet related target domain. The core idea of domain adaptation is to map data from different domains into the same space for comparison. Here, we apply two popular methods: the domain Discriminator network \cite{discriminator_da,wang2018deep,discriminator_da2} and the Multi-Kernel Maximum Mean Discrepancy (MK-MMD) function \cite{mkmmd1,mkmmd2,wang2018deep} to compare the feature spaces of different models.
\section{Method}
\label{sec:Method}

\begin{figure}[t]
    \centering
    \subfloat[Original]{\includegraphics[width=0.70in]{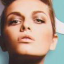}}
    \hfill
    \subfloat[Model 1]{\includegraphics[width=0.70in]{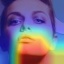}}
    \hfill
    \subfloat[Model 2]{\includegraphics[width=0.70in]{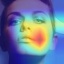}}
    \hfill
    \subfloat[Model 3]{\includegraphics[width=0.70in]{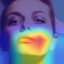}}
    \\
    % \subfloat[Original]{\includegraphics[width=0.70in]{img/cam/original2.png}}
    % \hfill
    % \subfloat[Victim]{\includegraphics[width=0.70in]{img/cam/target_19.jpg}}
    % \hfill
    % \subfloat[FORA]{\includegraphics[width=0.70in]{img/cam/pseudo_19.jpg}}
    % \hfill
    % \subfloat[PCAT]{\includegraphics[width=0.70in]{img/cam/pcat_19.jpg}}
\caption{Input image and behavior visualization by Grad-CAM \cite{selvaraju2017grad}. All the models are trained in CelebA with the task of smiling classification. The figure displays the original images and the representation preferences of three models trained under the same hyperparameter settings but with different random seeds.}
% The second row displays the original images and the representation preferences of the victim client and substitute clients of FORA and PCAT.
\label{fig:gradcam}
\end{figure}

\begin{figure*}[t]
\centering
\includegraphics[width=6.0in]{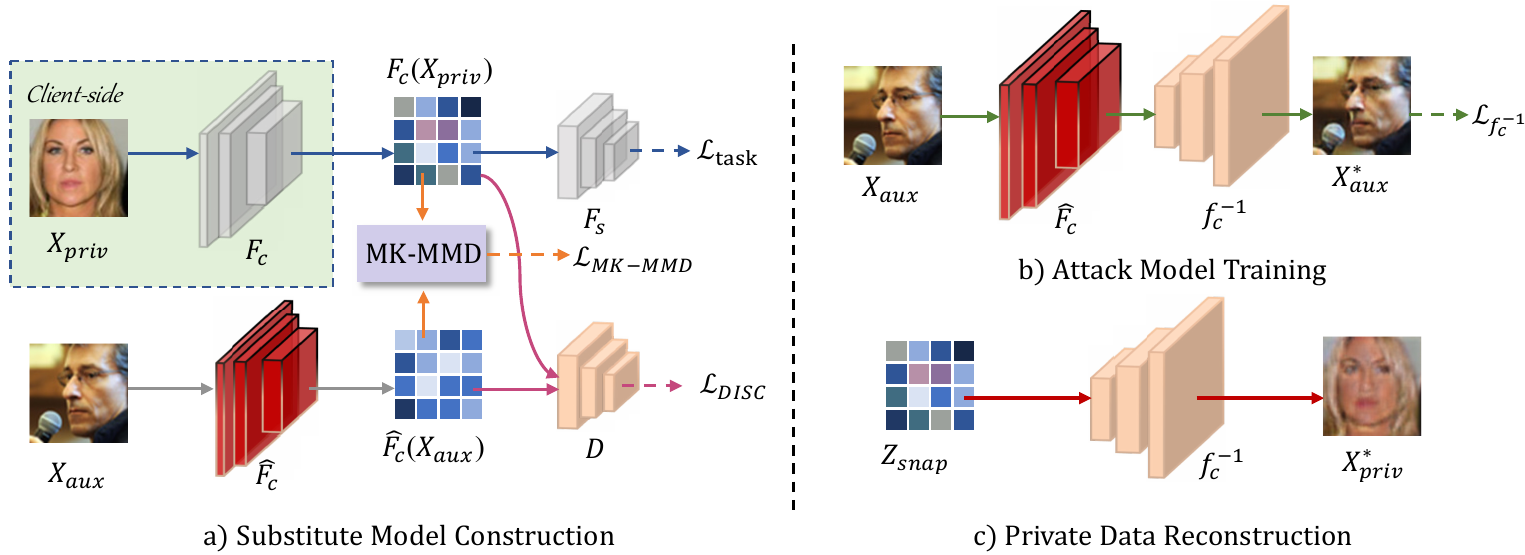}
\caption{Attack pipeline of Feature-Oriented Reconstruction
Attack (FORA) against SL. (a) shows the substitute model training phase. The attacker constructs a substitute model $\hat{F_c}$ using $\mathcal{L}_{DISC}$ and $\mathcal{L}_{MK-MMD}$ to mimic the behavior of the client model $F_c$. (b) means training an inverse network $f^{-1}_{c}$   using public data $X_{aux}$. (c) represents the final attack phase using the attack model to reconstruct training data from snapshot $Z_{snap}$ of target smashed data.}
\label{fig:method_share}
\end{figure*}

% In this section, we first discuss a more practical threat model for the SL system. Then, we explore the inherent privacy leakage of SL and provide our new insights. Finally, we present our proposed attack scheme in detail.

% previous version
% In this section, we will first discuss our threat model based on more realistic assumptions. Then following our main intuition, we propose a new stealing attack, called Feature-Oriented Stealing Attacks (FOSA). 
% In this section,  we first present our new insights on how to exploit the inherent vulnerabilities of SL. Then, we introduced our setting against more realistic SL systems. Finally, we elaborate on the details about our proposed FORA.

\subsection{Threat Model}
\label{subsec:Threat Model}

% \textbf {Adversary’s Goal.}
Without loss of generality, given a two-party SL protocol, the SL model $F$ is partitioned to a server model $F_s$ and a client model $F_c$. The server aims to stealthily recover the private training data of the client through the smashed data $Z$ output by $F_c$.

% previous version
% Without loss of generality, given a two-party SL protocol, the model $F$ is partitioned to a server model $F_{s}$ and a client model $F_{c}$. And server aims to infer the raw data about the client model via smashed data $Z_c$ output by the $F_{c}$ secretly.

% \textbf {Adversary’s Capability.}
% We make the assumption that the server adversary operates as a semi-honest entity, ensuring that the training process, when performing the attack, remains indistinguishable from regular training in the client's perspective. 
We assume that the server adversary is a semi-honest entity, ensuring that the training process is indistinguishable from ordinary training during attack. 
Furthermore, we posit that the server adversary must adhere to the foundational principle of the SL — she lacks any means of accessing client-sensitive information. Specifically, the server does not require knowledge of the structure or hyperparameters of $F_c$ and is devoid of access to the client's private training dataset $D_{priv}$. The sole piece of public knowledge available to the server pertains to the auxiliary dataset $D_{aux}$, sourced from the same domain as the private samples. It's important to note that the distribution of $D_{aux}$ typically differs from that of $D_{priv}$. 
Compared to the threat model of previous works, this assumption is more reasonable and realistic.

\subsection{Motivation}
\label{subsec:New Insights}
Current DRAs rely overly on constructing inverse networks from input-output pairs obtained by querying the target model. However, this approach is impractical for SL because the server only has access to the client's outputs and is not qualified to query. A potential solution is to build a substitute client to mimic the target client, thus enabling the training of the inverse network. However, the variability of the substitute client's behavior affects the generalization of the inverse network to the target client, leading to the failure of the reconstruction, especially without the knowledge of the client model structure and private data distribution.

% previous version
% Current DRA often rely heavily on input-output pairs obtained by querying the target model to train an inverse network. However, this approach is impractical against SL as the server only has the output of client  without raw data.  Tackling the trials, a potential solution is to construct a substitute client that mimics the behavior of the target client, enabling the training of an inverse network. Nevertheless, there exists a significant challenge for the server since the variability in client behaviors can impact the generalization of the inverse network for target client, leading to unreasonable reconstructed images, particularly when unaware of the client's architecture and the distribution of the private dataset.

As illustrated in \cref{fig:gradcam}, we employ Grad-CAM \cite{selvaraju2017grad} to visualize the attention of intermediate features generated by different clients. From \cref{fig:gradcam} (a)-(d), it can be noticed that even for models trained under the same setup, there still exists evident differences between their image processing attention. This phenomenon suggests that the smashed data output by the client reflects its distinctive feature extraction behavior, which we define as representation preferences. Our general assumption is that narrowing the gap between the substitute client and the target client in terms of intermediate features can make the representation preferences of the two models more similar, which ensures that the inverse network trained by the substitute client perfectly maps the target smashed data back to the private raw data. 

\subsection{Feature-Oriented Reconstruction Attack}
\label{subsec:Feature-Oriented Reconstruction Attack}
Inspired by the differences in model representation preferences, we propose a novel data reconstruction attack against SL, called Feature-Oriented Reconstruction Attack (FORA). In order to mount FORA, the adversary needs to contrive a way to obtain the representation preferences of the $F_c$. To address this problem, we utilize domain adaptation techniques \cite{mkmmd1,mkmmd2,discriminator_da} to project different preference representations into the same space. Specifically, the adversary conducts feature-level transfer learning by exploiting the $Z_c$ collected in each training iteration and then obtains a substitute model that mimics well the feature extraction behavior of the $F_c$. Through this approach, the adversary can smoothly construct an attack model (inverse mapping network) to recover the private samples. The detailed pipeline of FORA is shown in \cref{fig:method_share}. It consists of three phases: substitute model construction, attack model training, and private data reconstruction.

% previous version
% Later, we will provide a detailed explanation of our attack mechanism Feature-Oriented Reconstruction Attack (FORA), which consists of two steps. In each iteration, the server initially trains a substitute client and subsequently constructs an inverse network to reconstruct the training set. Our attack pipeline are illustrated in \cref{fig:method_share}.

\textbf{Substitute Model Construction.}
Before SL training commences, the server initializes a substitute client, denoted by $\hat{F_c}$. The $\hat{F_c}$ will be trained locally at the server in parallel with the victim's $F_c$, and such process will take place throughout the entire SL collaboration. In each training iteration, the client will send smashed data of the current batch to the server for completing the subsequent computations. Concurrently, the server will use the collected smashed data to perform training on the $\hat{F_c}$. For this purpose, the server introduces the Discriminator module and the MK-MMD module to extract the representation preferences. We define its training objective as:
\begin{equation}
\label{eq:total}
 \mathop{\min}_{\hat{F_{c}}} \:  \mathcal{L}_{DISC} + \mathcal{L}_{MK-MMD},
\end{equation}
% \begin{equation}
% \label{eq:total}
%  \mathcal{L}_{\hat{F_{c}}} = \mathcal{L}_{DISC} + \mathcal{L}_{MK-MMD},
% \end{equation}
where $\mathcal{L}_{DISC}$ is the Discriminator module constraining $Z_{aux} = \hat{F_c}(X_{aux})$ and $Z_{priv} = F_c(X_{priv})$ to be indistinguishable, while $\mathcal{L}_{MK-MMD}$ is the MK-MMD module making $Z_{aux}$ as close as possible to $Z_{priv}$ in shared space.

The Discriminator \cite{gan,wgan,discriminator_da} $D$ is also a network that needs to be trained synchronously and is tasked with efficiently distinguishing the generated features between $F_{c}$ and $\hat{F_{c}}$, maximizing probabilities of the former and minimizing probabilities of the latter \cite{pasquini2021unleashing}. Therefore, the parameters of $D$ will be updated to minimize the following loss function:
% \begin{equation}
% \mathcal{L}_{D} = \log{(1-\mathcal{D}(F_{c}(X_{aux}))} + \log{\mathcal{D}(\hat{F_{c}}(X_{priv}))}.
% \end{equation}
\begin{equation}
\mathcal{L}_{D} = \log{(1-\mathcal{D}(F_{c}(X_{priv}))} + \log{\mathcal{D}(\hat{F_{c}}(X_{aux}))}.
\end{equation}
After each local training step of $D$, the server utilizes $D$ to instruct substitute client's representation preference to be consistent with that of the victim client. Specifically, an adversarial loss is constructed as the following:
% the loss function is constructed   from an adversarial perspective:
\begin{equation}
\mathcal{L}_{DISC} = \log{(1-D(\hat{F_c}(X_{aux})))}.
\end{equation}

The MK-MMD module \cite{mkmmd1,mkmmd2} is designed to align two sets of generated features into a shared space using kernel functions and compute their difference, where a smaller difference signifies closer representation preferences. Then, for the substitute client, the objective extends beyond maximizing the probabilities output by the $D$, it also seeks to minimize the MK-MMD loss function, namely:
\begin{gather}
\mathcal L_{MK-MMD} =\left\|\phi\left(\hat{F_c}(X_{aux})\right)- \phi\left(F_{c}(X_{priv})\right)\right\|_{\mathcal{H}}, \\
\left\{     
    \begin{aligned}     
        \mathcal{\phi} & = \sum_{j=1}^m \beta_j k_j, \\ 
        \sum_{j=1}^m \beta_j & = 1, \beta_j \geq0, \forall j,
    \end{aligned}
\right.
\end{gather}
where $k$ is a single kernel function, $\phi$ denotes a set of kernel functions that project different smashed data into  Reproducing Kernel Hilbert Space $\mathcal{H}$, $\beta$ is the weight coefficient corresponding to the single kernel function.

\textbf{Attack Model Training.}
At the end of the training of SL, the server can obtain a substitute client with a feature extraction behavior extremely similar to that of the victim client.
Moreover, its feature space is known to the adversary, who can recover the original input from the smashed data by applying an inverse network (denoted as $f_c^{-1}$).
Following previous DRAs \cite{he2019model,disco}, we adopt the $f_c^{-1}$ consisting of a set of Transposed Convolution layers and Tanh activations as our attack model. The server can leverage the auxiliary dataset to train the attack model by minimizing the mean square error between $f_c^{-1}(\hat{F_c}(X_{aux}))$ and $X_{aux}$ as follows:
\begin{equation}
\mathcal L_{f_c^{-1}} =\|f_c^{-1}(\hat{F_c}(X_{aux})) - X_{aux}\|_2^2.
\end{equation}

\textbf{Private Data Reconstruction.}
% The server keeps a snapshot $F_c(X_{priv})$ of all smashed data output by the target client under the final training iteration for reconstruction, where $F^{N}_{c}$ represents the client that has undergone N rounds of training.
The server keeps a snapshot $Z_{snap}=F_c(X_{priv})$ of all smashed data output by the target client under the final training iteration for reconstruction. Since the substitute client is able to mimic the target client's representation preferences well, the server can subtly use $f_c^{-1}$ to perform the attack by mapping the target smashed data directly into the private raw data space, namely:
\begin{equation}
X_{priv}^* = f_c^{-1}(Z_{snap}).
\end{equation}
Here, $X_{priv}^*$ are the reconstructed private training samples.

\section{Experiments}
\label{sec:Experiments}
% In this section, we describe our experimental setup and conduct comprehensive experiments to demonstrate the attack effectiveness and broad applicability of FORA.

% In this section, we first describe our experimental setup, including the task dataset, model architecture, and attack baseline. Then, we conduct comprehensive experiments to demonstrate the attack effectiveness and broad applicability of the proposed FORA.

% previous version
% In this section, we first introduce our experimental setup, including model architectures, dataset partitioning, and the baselines we compare against. Then we conduct comprehensive experiments to thoroughly evaluate the effectiveness of the proposed FOSA.

\subsection{Experimental Setup} 
\label{subsec:Experimental Setup}

\textbf{Datasets.}
In our experiments, we rely on CIFAR-10 \cite{cifar} and CelebA \cite{celeba} to validate the attacks, due to their dominance in the research on SL \cite{pasquini2021unleashing, gaopcat, erdougan2022unsplit}. They will be used as private data for the client's target training tasks.
According to \cref{subsec:Threat Model}, we assume that the server adversary has access to a set of auxiliary samples that are distinct from the client's private data. Therefore, we choose CINIC-10 \cite{cinic} and FFHQ \cite{ffhq} as the adversary's auxiliary dataset, respectively. We exclude images in CINIC-10 that overlapped with CIFAR-10, and randomly select 5,000 samples and 10,000 samples from the preprocessed CINIC-10 and FFHQ as the final auxiliary data.
\cref{appendix_dataset} provides the detailed information for different datasets. 

\textbf{Models.}
We consider two popular types of neural network architectures, including MobileNet \cite{howard2017mobilenets} and ResNet-18 \cite{he2016deep}, as target models for the classification tasks of CIFRA-10 and CelebA, respectively.
% Different splitting points are set for the target models to show the results of the attack methods when dealing with various network depths. 
We set various split points for different target models to show our attack performance.
Since the server is entirely unaware of the client's model structure from \cref{subsec:Threat Model}, we use VGG blocks \cite{vgg} (consisting of a sequence of Convolutional, BatchNorm, ReLU, and MaxPool layers) to construct substitute models. In addition, the adversary's substitute models adaptively depend on the size of the intermediate features output by the client. All the architecture information and splitting schemes used in this paper are reported in \cref{appendix_arc}.

% previous version
% We evaluated our method using two popular target models: MobileNet \cite{howard2017mobilenets} and ResNet-18 \cite{he2016deep}. These two networks correspond to the classification tasks on the  CIFAR-10, and CelebA respectively. For each network, we set different splitting points to demonstrate the results of our approach when dealing with different network depths. The target network splitting schemes are illustrated in \cref{fig:model_split}. We assume that the server is entirely unaware of the client's architecture, so, following our assumptions, we chooses VGG blocks to construct substitute model. Furthermore, regardless of the dataset, classification task, or the architecture of the target client model, our substitute model is solely determined by the feature size output by the client. For all target clients with the same output feature size, the server adopts the same substitute model architecture. Unless specified otherwise in subsequent experiments, we consistently adopt this configuration. All employed architectures in this paper are reported in  \cref{tab:model_details} at \cref{appendix_other_arc}.

\textbf{Metrics.}
In addition to analyzing the qualitative results of attack performances visually, we chose three quantitative metrics to evaluate the quality of the reconstructed images: Structural Similarity (SSIM) \cite{ssim}, Peak Signal-to-Noise Ratio (PSNR) \cite{psnr}, and Learned Perceptual Image Patch Similarity (LPIPS) \cite{lpips}. We also use Cosine Similarity and Mean Square Error to measure the similarity between the substitute client and the target client in feature space.

% previous version
% We use  cosine similarity and  mean squared error to measure the similarity in the feature space between the substitite client and the target client. In addition to visually quantifying attack performance, we selected three metrics to evaluate the quality of the reconstructed images:  structural similarity (SSIM) \cite{ssim},  Peak Signal-to-Noise Ratio (PSNR) \cite{psnr},  Learned Perceptual Image Patch Similarity (LPIPS) \cite{lpips}.

\textbf{Attack Baselines.}
We mainly compare our approach with three representative existing methods, which are FSHA \cite{pasquini2021unleashing}, UnSplit \cite{erdougan2022unsplit}, and PCAT \cite{gaopcat}. For the malicious attack FSHA, we use sophisticated detection mechanism to jointly evaluate the attack's effectiveness. For the semi-honest attack UnSplit, we make it consistent with our experimental settings to ensure fairness. PCAT requires an understanding of the learning task while relying on a subset of the private training data to build the pseudo-client, and in order to comply with this assumption, we set the proportion of the CIFAR-10 private dataset to be 5\% (the maximal threshold suggested by the original paper), and for more complex CelebA dataset, we extend the proportion to be 10\%.

% previous version
% In this work, we consider our method with the three attacks: FSHA \cite{pasquini2021unleashing}, PCAT \cite{gaopcat} and UnSplit \cite{erdougan2022UnSplit}. To ensure fairness in the experimental results, we use the same model architecture when comparing with the baselines. For FSHA and UnSplit,  we use the same dataset settings as ours. It is worth noting that we are the first work to perform passive attacks on SL under such stringent assumptions, requiring only an unlabeled public dataset. PCAT requires the knowledge of the learning task and relies on a subset of private data to train a pseudo model. In adherence to the original paper's guidelines, we utilized \textbf{a part of privacy dataset extracted from the target training se for PCAT.} The proportion of the CIFAR-10 private dataset remains consistent with the original paper at 5\%. For the more complex face dataset, we set the proportion at 10\%, which exceeds the maximum threshold 5\% recommended in the paper.

\begin{table*}[t]
\centering
\caption{Data reconstruction results of UnSplit, PCAT, and FORA on CIFAR-10 and CelebA in different splitting settings.}
\tabcolsep=1pt
\resizebox{\linewidth}{!}{
\begin{tabular}{c|@{\hspace{2pt}}c@{\hspace{0pt}}c@{\hspace{0pt}}c@{\hspace{2pt}}
|@{\hspace{2pt}}c@{\hspace{0pt}}c@{\hspace{0pt}}c@{\hspace{2pt}}
|@{\hspace{2pt}}c@{\hspace{0pt}}c@{\hspace{0pt}}c@{\hspace{2pt}}
|@{\hspace{2pt}}c@{\hspace{0pt}}c@{\hspace{0pt}}c@{\hspace{2pt}}
|@{\hspace{2pt}}c@{\hspace{0pt}}c@{\hspace{0pt}}c@{\hspace{2pt}}
|@{\hspace{2pt}}c@{\hspace{0pt}}c@{\hspace{0pt}}c@{\hspace{2pt}}}
    \hline
    \multirow{2}{*}{\footnotesize{\makecell{Split \\Point}}} & \multicolumn{9}{c|@{\hspace{2pt}}}{\footnotesize{CIFAR-10}} & \multicolumn{9}{c}{\footnotesize{CelebA}} \vspace{-1mm}\\
    % \cline{2-9}
    % \cline{12-19}
    % \vspace{-1.5mm}
    & \multicolumn{3}{c}{\footnotesize{UnSplit}} & \multicolumn{3}{c}{\footnotesize{PCAT}} & \multicolumn{3}{c|@{\hspace{2pt}}}{\footnotesize{FORA}}
    & \multicolumn{3}{c}{\footnotesize{UnSplit}} & \multicolumn{3}{c}{\footnotesize{PCAT}} & \multicolumn{3}{c}{\footnotesize{FORA}}\\
    \hline
    \footnotesize{\makecell{Ground \\Truth} }
    & \begin{minipage}[c]{0.10\columnwidth}
    \centering
    {\includegraphics[width=1.0\textwidth]{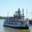}}
    \end{minipage}
    & \begin{minipage}[c]{0.10\columnwidth}
    \centering
    {\includegraphics[width=1.0\textwidth]{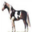}}
    \end{minipage}
    & \begin{minipage}[c]{0.10\columnwidth}
    \centering
    {\includegraphics[width=1.0\textwidth]{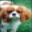}}
    \end{minipage}
    & \begin{minipage}[c]{0.10\columnwidth}
    \centering
    {\includegraphics[width=1.0\textwidth]{img/evaluation/baselines/ship_truth.png}}
    \end{minipage}
    & \begin{minipage}[c]{0.10\columnwidth}
    \centering
    {\includegraphics[width=1.0\textwidth]{img/evaluation/baselines/horse_truth.png}}
    \end{minipage}
    & \begin{minipage}[c]{0.10\columnwidth}
    \centering
    {\includegraphics[width=1.0\textwidth]{img/evaluation/baselines/dog_truth.png}}
    \end{minipage}
    & \begin{minipage}[c]{0.10\columnwidth}
    \centering
    {\includegraphics[width=1.0\textwidth]{img/evaluation/baselines/ship_truth.png}}
    \end{minipage}
    & \begin{minipage}[c]{0.10\columnwidth}
    \centering
    {\includegraphics[width=1.0\textwidth]{img/evaluation/baselines/horse_truth.png}}
    \end{minipage}
    & \begin{minipage}[c]{0.10\columnwidth}
    \centering
    {\includegraphics[width=1.0\textwidth]{img/evaluation/baselines/dog_truth.png}}
    \end{minipage}
    & \begin{minipage}[c]{0.10\columnwidth}
    \centering
    {\includegraphics[width=1.0\textwidth]{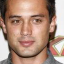}}
    \end{minipage}
    & \begin{minipage}[c]{0.10\columnwidth}
    \centering
    {\includegraphics[width=1.0\textwidth]{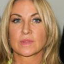}}
    \end{minipage}
    & \begin{minipage}[c]{0.10\columnwidth}
    \centering
    {\includegraphics[width=1.0\textwidth]{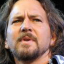}}
    \end{minipage}
    & \begin{minipage}[c]{0.10\columnwidth}
    \centering
    {\includegraphics[width=1.0\textwidth]{img/evaluation/baselines/face1_truth.png}}
    \end{minipage}
    & \begin{minipage}[c]{0.10\columnwidth}
    \centering
    {\includegraphics[width=1.0\textwidth]{img/evaluation/baselines/face2_truth.png}}
    \end{minipage}
    & \begin{minipage}[c]{0.10\columnwidth}
    \centering
    {\includegraphics[width=1.0\textwidth]{img/evaluation/baselines/face3_truth.png}}
    \end{minipage}
    & \begin{minipage}[c]{0.10\columnwidth}
    \centering
    {\includegraphics[width=1.0\textwidth]{img/evaluation/baselines/face1_truth.png}}
    \end{minipage}
    & \begin{minipage}[c]{0.10\columnwidth}
    \centering
    {\includegraphics[width=1.0\textwidth]{img/evaluation/baselines/face2_truth.png}}
    \end{minipage}
    & \begin{minipage}[c]{0.10\columnwidth}
    \centering
    {\includegraphics[width=1.0\textwidth]{img/evaluation/baselines/face3_truth.png}}
    \end{minipage}
    \\
    \hline
    \footnotesize{layer 1 }
    & \begin{minipage}[c]{0.10\columnwidth}
    \centering
    {\includegraphics[width=1.0\textwidth]{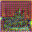}}
    \end{minipage}
    & \begin{minipage}[c]{0.10\columnwidth}
    \centering
    {\includegraphics[width=1.0\textwidth]{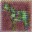}}
    \end{minipage}
    & \begin{minipage}[c]{0.10\columnwidth}
    \centering
    {\includegraphics[width=1.0\textwidth]{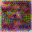}}
    \end{minipage}
    & \begin{minipage}[c]{0.10\columnwidth}
    \centering
    {\includegraphics[width=1.0\textwidth]{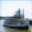}}
    \end{minipage}
    & \begin{minipage}[c]{0.10\columnwidth}
    \centering
    {\includegraphics[width=1.0\textwidth]{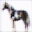}}
    \end{minipage}
    & \begin{minipage}[c]{0.10\columnwidth}
    \centering
    {\includegraphics[width=1.0\textwidth]{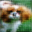}}
    \end{minipage}
    & \begin{minipage}[c]{0.10\columnwidth}
    \centering
    {\includegraphics[width=1.0\textwidth]{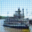}}
    \end{minipage}
    & \begin{minipage}[c]{0.10\columnwidth}
    \centering
    {\includegraphics[width=1.0\textwidth]{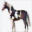}}
    \end{minipage}
    & \begin{minipage}[c]{0.10\columnwidth}
    \centering
    {\includegraphics[width=1.0\textwidth]{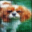}}
    \end{minipage}
    & \begin{minipage}[c]{0.10\columnwidth}
    \centering
    {\includegraphics[width=1.0\textwidth]{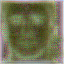}}
    \end{minipage}
    & \begin{minipage}[c]{0.10\columnwidth}
    \centering
    {\includegraphics[width=1.0\textwidth]{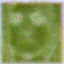}}
    \end{minipage}
    & \begin{minipage}[c]{0.10\columnwidth}
    \centering
    {\includegraphics[width=1.0\textwidth]{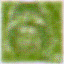}}
    \end{minipage}
    & \begin{minipage}[c]{0.10\columnwidth}
    \centering
    {\includegraphics[width=1.0\textwidth]{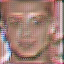}}
    \end{minipage}
    & \begin{minipage}[c]{0.10\columnwidth}
    \centering
    {\includegraphics[width=1.0\textwidth]{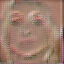}}
    \end{minipage}
    & \begin{minipage}[c]{0.10\columnwidth}
    \centering
    {\includegraphics[width=1.0\textwidth]{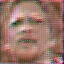}}
    \end{minipage}
    & \begin{minipage}[c]{0.10\columnwidth}
    \centering
    {\includegraphics[width=1.0\textwidth]{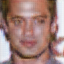}}
    \end{minipage}
    & \begin{minipage}[c]{0.10\columnwidth}
    \centering
    {\includegraphics[width=1.0\textwidth]{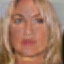}}
    \end{minipage}
    & \begin{minipage}[c]{0.10\columnwidth}
    \centering
    {\includegraphics[width=1.0\textwidth]{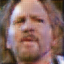}}
    \end{minipage}
    \\
    \hline
    \footnotesize{layer 2}
    & \begin{minipage}[c]{0.10\columnwidth}
    \centering
    {\includegraphics[width=1.0\textwidth]{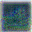}}
    \end{minipage}
    & \begin{minipage}[c]{0.10\columnwidth}
    \centering
    {\includegraphics[width=1.0\textwidth]{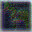}}
    \end{minipage}
    & \begin{minipage}[c]{0.10\columnwidth}
    \centering
    {\includegraphics[width=1.0\textwidth]{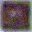}}
    \end{minipage}
    & \begin{minipage}[c]{0.10\columnwidth}
    \centering
    {\includegraphics[width=1.0\textwidth]{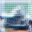}}
    \end{minipage}
    & \begin{minipage}[c]{0.10\columnwidth}
    \centering
    {\includegraphics[width=1.0\textwidth]{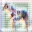}}
    \end{minipage}
    & \begin{minipage}[c]{0.10\columnwidth}
    \centering
    {\includegraphics[width=1.0\textwidth]{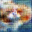}}
    \end{minipage}
    & \begin{minipage}[c]{0.10\columnwidth}
    \centering
    {\includegraphics[width=1.0\textwidth]{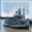}}
    \end{minipage}
    & \begin{minipage}[c]{0.10\columnwidth}
    \centering
    {\includegraphics[width=1.0\textwidth]{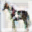}}
    \end{minipage}
    & \begin{minipage}[c]{0.10\columnwidth}
    \centering
    {\includegraphics[width=1.0\textwidth]{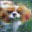}}
    \end{minipage}
    & \begin{minipage}[c]{0.10\columnwidth}
    \centering
    {\includegraphics[width=1.0\textwidth]{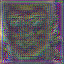}}
    \end{minipage}
    & \begin{minipage}[c]{0.10\columnwidth}
    \centering
    {\includegraphics[width=1.0\textwidth]{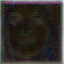}}
    \end{minipage}
    & \begin{minipage}[c]{0.10\columnwidth}
    \centering
    {\includegraphics[width=1.0\textwidth]{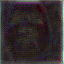}}
    \end{minipage}
    & \begin{minipage}[c]{0.10\columnwidth}
    \centering
    {\includegraphics[width=1.0\textwidth]{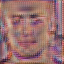}}
    \end{minipage}
    & \begin{minipage}[c]{0.10\columnwidth}
    \centering
    {\includegraphics[width=1.0\textwidth]{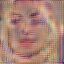}}
    \end{minipage}
    & \begin{minipage}[c]{0.10\columnwidth}
    \centering
    {\includegraphics[width=1.0\textwidth]{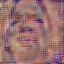}}
    \end{minipage}
    & \begin{minipage}[c]{0.10\columnwidth}
    \centering
    {\includegraphics[width=1.0\textwidth]{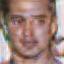}}
    \end{minipage}
    & \begin{minipage}[c]{0.10\columnwidth}
    \centering
    {\includegraphics[width=1.0\textwidth]{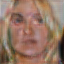}}
    \end{minipage}
    & \begin{minipage}[c]{0.10\columnwidth}
    \centering
    {\includegraphics[width=1.0\textwidth]{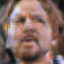}}
    \end{minipage}
    \\
    \hline
    \footnotesize{layer 3 }
    & \begin{minipage}[c]{0.10\columnwidth}
    \centering
    {\includegraphics[width=1.0\textwidth]{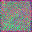}}
    \end{minipage}
    & \begin{minipage}[c]{0.10\columnwidth}
    \centering
    {\includegraphics[width=1.0\textwidth]{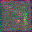}}
    \end{minipage}
    & \begin{minipage}[c]{0.10\columnwidth}
    \centering
    {\includegraphics[width=1.0\textwidth]{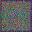}}
    \end{minipage}
    & \begin{minipage}[c]{0.10\columnwidth}
    \centering
    {\includegraphics[width=1.0\textwidth]{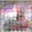}}
    \end{minipage}
    & \begin{minipage}[c]{0.10\columnwidth}
    \centering
    {\includegraphics[width=1.0\textwidth]{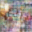}}
    \end{minipage}
    & \begin{minipage}[c]{0.10\columnwidth}
    \centering
    {\includegraphics[width=1.0\textwidth]{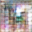}}
    \end{minipage}
    & \begin{minipage}[c]{0.10\columnwidth}
    \centering
    {\includegraphics[width=1.0\textwidth]{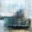}}
    \end{minipage}
    & \begin{minipage}[c]{0.10\columnwidth}
    \centering
    {\includegraphics[width=1.0\textwidth]{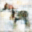}}
    \end{minipage}
    & \begin{minipage}[c]{0.10\columnwidth}
    \centering
    {\includegraphics[width=1.0\textwidth]{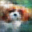}}
    \end{minipage}
    & \begin{minipage}[c]{0.10\columnwidth}
    \centering
    {\includegraphics[width=1.0\textwidth]{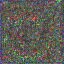}}
    \end{minipage}
    & \begin{minipage}[c]{0.10\columnwidth}
    \centering
    {\includegraphics[width=1.0\textwidth]{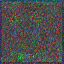}}
    \end{minipage}
    & \begin{minipage}[c]{0.10\columnwidth}
    \centering
    {\includegraphics[width=1.0\textwidth]{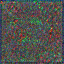}}
    \end{minipage}
    & \begin{minipage}[c]{0.10\columnwidth}
    \centering
    {\includegraphics[width=1.0\textwidth]{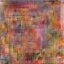}}
    \end{minipage}
    & \begin{minipage}[c]{0.10\columnwidth}
    \centering
    {\includegraphics[width=1.0\textwidth]{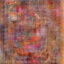}}
    \end{minipage}
    & \begin{minipage}[c]{0.10\columnwidth}
    \centering
    {\includegraphics[width=1.0\textwidth]{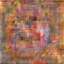}}
    \end{minipage}
    & \begin{minipage}[c]{0.10\columnwidth}
    \centering
    {\includegraphics[width=1.0\textwidth]{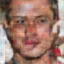}}
    \end{minipage}
    & \begin{minipage}[c]{0.10\columnwidth}
    \centering
    {\includegraphics[width=1.0\textwidth]{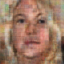}}
    \end{minipage}
    & \begin{minipage}[c]{0.10\columnwidth}
    \centering
    {\includegraphics[width=1.0\textwidth]{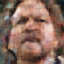}}
    \end{minipage}
    \\
    \hline
    \footnotesize{layer 4}
    & \begin{minipage}[c]{0.10\columnwidth}
    \centering
    {\includegraphics[width=1.0\textwidth]{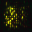}}
    \end{minipage}
    & \begin{minipage}[c]{0.10\columnwidth}
    \centering
    {\includegraphics[width=1.0\textwidth]{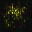}}
    \end{minipage}
    & \begin{minipage}[c]{0.10\columnwidth}
    \centering
    {\includegraphics[width=1.0\textwidth]{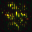}}
    \end{minipage}
    & \begin{minipage}[c]{0.10\columnwidth}
    \centering
    {\includegraphics[width=1.0\textwidth]{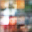}}
    \end{minipage}
    & \begin{minipage}[c]{0.10\columnwidth}
    \centering
    {\includegraphics[width=1.0\textwidth]{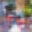}}
    \end{minipage}
    & \begin{minipage}[c]{0.10\columnwidth}
    \centering
    {\includegraphics[width=1.0\textwidth]{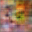}}
    \end{minipage}
    & \begin{minipage}[c]{0.10\columnwidth}
    \centering
    {\includegraphics[width=1.0\textwidth]{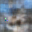}}
    \end{minipage}
    & \begin{minipage}[c]{0.10\columnwidth}
    \centering
    {\includegraphics[width=1.0\textwidth]{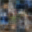}}
    \end{minipage}
    & \begin{minipage}[c]{0.10\columnwidth}
    \centering
    {\includegraphics[width=1.0\textwidth]{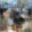}}
    \end{minipage}
    & \begin{minipage}[c]{0.10\columnwidth}
    \centering
    {\includegraphics[width=1.0\textwidth]{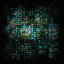}}
    \end{minipage}
    & \begin{minipage}[c]{0.10\columnwidth}
    \centering
    {\includegraphics[width=1.0\textwidth]{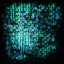}}
    \end{minipage}
    & \begin{minipage}[c]{0.10\columnwidth}
    \centering
    {\includegraphics[width=1.0\textwidth]{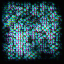}}
    \end{minipage}
    & \begin{minipage}[c]{0.10\columnwidth}
    \centering
    {\includegraphics[width=1.0\textwidth]{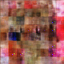}}
    \end{minipage}
    & \begin{minipage}[c]{0.10\columnwidth}
    \centering
    {\includegraphics[width=1.0\textwidth]{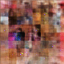}}
    \end{minipage}
    & \begin{minipage}[c]{0.10\columnwidth}
    \centering
    {\includegraphics[width=1.0\textwidth]{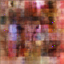}}
    \end{minipage}
    & \begin{minipage}[c]{0.10\columnwidth}
    \centering
    {\includegraphics[width=1.0\textwidth]{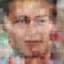}}
    \end{minipage}
    & \begin{minipage}[c]{0.10\columnwidth}
    \centering
    {\includegraphics[width=1.0\textwidth]{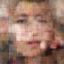}}
    \end{minipage}
    & \begin{minipage}[c]{0.10\columnwidth}
    \centering
    {\includegraphics[width=1.0\textwidth]{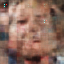}}
    \end{minipage}
    \\
    \hline
\end{tabular}
}
\label{tab:dra}
\end{table*}
%--------------------------------------------------------------
%---------------------------------------------------------
\begin{figure}[t]
\centering
  \subfloat[Detection Score]{
    \begin{minipage}[b]{0.55\columnwidth}
      \centering
      \includegraphics[width=1.8in]{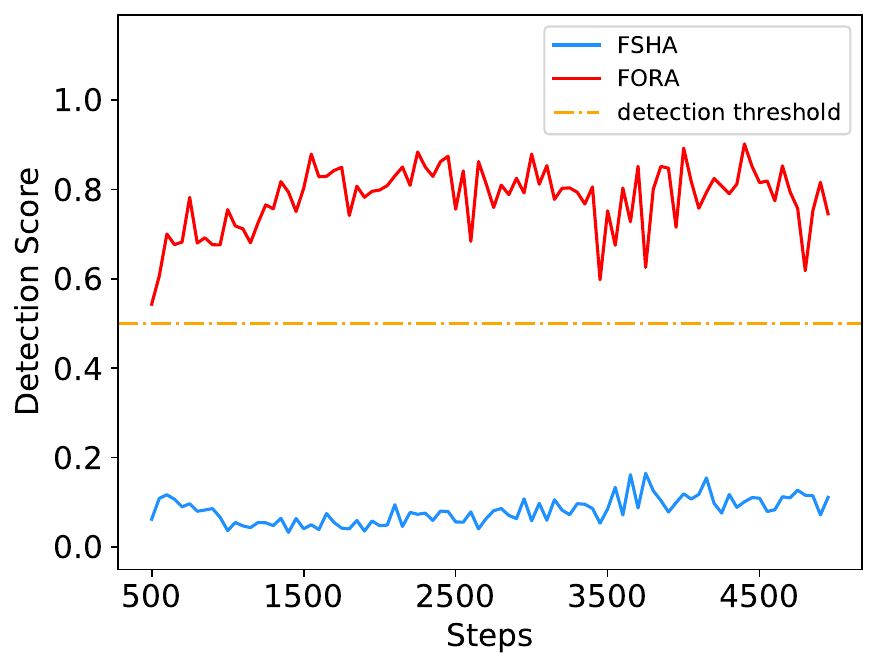}
      % \vspace{-1em}
    \end{minipage}
  }
  \hfill
  \subfloat[Reconstruction Results]{
    \begin{minipage}[b]{0.4\columnwidth}
      \centering
      \resizebox{\linewidth}{!}{
      \begin{tabular}{cccc}
        \huge{Truth} & \huge{FORA} & \huge{FSHA} & \huge{FSHA-GS} \\
        \begin{minipage}[c]{0.7\columnwidth}
        {\includegraphics[width=1\columnwidth]{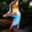}}
        \end{minipage}
        & \begin{minipage}[c]{0.7\columnwidth}
        {\includegraphics[width=1\columnwidth]{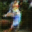}}
        \end{minipage}
        & \begin{minipage}[c]{0.7\columnwidth}
        {\includegraphics[width=1\columnwidth]{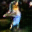}}
        \end{minipage}
        & \begin{minipage}[c]{0.7\columnwidth}
        {\includegraphics[width=1\columnwidth]{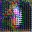}}
        \end{minipage}
        \\
        \begin{minipage}[c]{0.7\columnwidth}
        {\includegraphics[width=1\columnwidth]{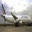}}
        \end{minipage}
        & \begin{minipage}[c]{0.7\columnwidth}
        {\includegraphics[width=1\columnwidth]{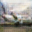}}
        \end{minipage}
        & \begin{minipage}[c]{0.7\columnwidth}
        {\includegraphics[width=1\columnwidth]{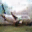}}
        \end{minipage}
        & \begin{minipage}[c]{0.7\columnwidth}
        {\includegraphics[width=1\columnwidth]{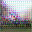}}
        \end{minipage}
        \\
        \begin{minipage}[c]{0.7\columnwidth}
        {\includegraphics[width=1\columnwidth]{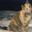}}
        \end{minipage}
        & \begin{minipage}[c]{0.7\columnwidth}
        {\includegraphics[width=1\columnwidth]{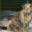}}
        \end{minipage}
        & \begin{minipage}[c]{0.7\columnwidth}
        {\includegraphics[width=1\columnwidth]{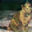}}
        \end{minipage}
        & \begin{minipage}[c]{0.7\columnwidth}
        {\includegraphics[width=1\columnwidth]{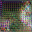}}
        \end{minipage}
        \\
        \begin{minipage}[c]{0.7\columnwidth}
        {\includegraphics[width=1\columnwidth]{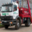}}
        \end{minipage}
        & \begin{minipage}[c]{0.7\columnwidth}
        {\includegraphics[width=1\columnwidth]{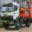}}
        \end{minipage}
        & \begin{minipage}[c]{0.7\columnwidth}
        {\includegraphics[width=1\columnwidth]{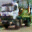}}
        \end{minipage}
        & \begin{minipage}[c]{0.7\columnwidth}
        {\includegraphics[width=1\columnwidth]{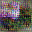}}
        \end{minipage}
        \\
        \begin{minipage}[c]{0.7\columnwidth}
        {\includegraphics[width=1\columnwidth]{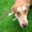}}
        \end{minipage}
        & \begin{minipage}[c]{0.7\columnwidth}
        {\includegraphics[width=1\columnwidth]{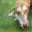}}
        \end{minipage}
        & \begin{minipage}[c]{0.7\columnwidth}
        {\includegraphics[width=1\columnwidth]{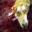}}
        \end{minipage}
        & \begin{minipage}[c]{0.7\columnwidth}
        {\includegraphics[width=1\columnwidth]{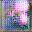}}
        \end{minipage}
        \\
      \end{tabular}
      }
      \vspace{0em}
    \end{minipage}
  }
  \caption{Attack performance  comparison of  FSHA \cite{pasquini2021unleashing} and FORA  on CIFAR-10 with layer 2. (a) shows the detection score of two attacks detected by GS. (b) represents the reconstruction results of two attacks, and FSHA-GS is the reconstructed images when detected by GS. }
  \label{fig:malicious}
\end{figure}
%----------------------------------------------------------

\subsection{Comparison with Malicious Attack}
\label{subsec:Comparison with Malicious Attack}
Since FSHA severely undermines the utility of the target client, recent work has proposed the Gradients Scrutinizer (GS) \cite{hust_ndss} to defend against such hijacking attacks by detecting the gradients returned from the server to the client. The GS will perform a similarity computation on the gradients, and if the calculated value is lower than a set threshold, it will be considered as a potential attack, resulting in the training of SL being immediately suspended. More details about GS can be found in \cref{appendix_defense_details}. We can observe from \cref{fig:malicious} that the reconstruction results of FORA are almost the same as those of FSHA in the unprotected SL system. Although FSHA performs well in capturing fine graphical details, it also leads to noticeable color shifts in some reconstruction results. Moreover, since FSHA drastically tampers with the updated gradient returned to the client model, it is easily detected by GS, leading to the failure of reconstruction.

% previous version
% Due to the significant disruption to client utility by FSHA, Gradients Scrutinizer (GS) \cite{hust_ndss} has been proposed for defense through detection the gradients returned from server. GS will perform similarity calculations on the returned gradients. If the computed values fall below the set threshold, it is considered a hijacking attack, resulting in an immediate halt to the SL. More details about GS are provided in \cref{appendix_defense}. 
% % Then we test the effectiveness of FSHA and our method on split2 before and after defense. The model configuration and dataset settings remain as described in \cref{setup}.

% \textbf{Experiment Results.} From \cref{fig:malicious}, we can observe that the reconstruction results of FSHA are almost consistent to ours. Although FSHA exhibits superior performance in capturing finer details, it can also lead to noticeable color shifts in some images. Moreover, due to the severe degradation of SL utility by FSHA, it is susceptible to be detected by GS, leading to poor results.

\subsection{Comparison with Semi-Honest Attacks}
\label{subsec:Comparison with Semi-Honest Attacks}

\textbf{Reconstruction Performance.}
We show in detail the reconstruction results for UnSplit, PCAT, and our proposed FORA on all split points for both datasets. As depicted in \cref{tab:dra}, compared to other attacks, the images reconstructed by FORA exhibit a significant improvement visually. Due to the vast search space and inefficient optimization approach, UnSplit almost fails to recover training data in both datasets, even at layer 1. Although PCAT can reconstruct training samples in the shallow settings of the CIFAR-10 dataset, such as layer 1 and layer 2, the reconstruction quality is still lower than that of FORA. For the more complex CelebA dataset, PCAT struggles to produce quality reconstructions. 
\cref{tab:dra_cifar} and \cref{tab:dra_celeba} provides the quantitative results of the attacks. Except for the anomaly at the layer 4 split point of CIFAR-10, where FORA slightly underperforms PCAT in terms of SSIM and PSNR metric, FORA is superior to both methods in all other settings, especially in terms of the LPIPS metric, which is considered to be more aligned with human perception. Notably, even though PCAT has access to a subset of the private data, while FORA only obtains samples with different distributions, FORA substantially surpasses PCAT for reconstruction. This further emphasizes the robust privacy threat our approach poses to SL.
More reconstructed images are presented in \cref{appendix_additional_results_semi}.

\begin{table}[tbp]
\centering
\caption{SSIM, PSNR, and LPIPS of the reconstructed images on CIFAR-10 among three attacks.}
\resizebox{\linewidth}{!}{
\begin{tabular}{c |c c c| ccc | ccc }
    \hline
    \multirow{2}{*}{\makecell{Split \\Point}} & \multicolumn{3}{c|}{SSIM$\uparrow$} & \multicolumn{3}{c|}{PSNR$\uparrow$} & \multicolumn{3}{c}{LPIPS$\downarrow$}\\
    & UnSplit & PCAT & FORA & UnSplit & PCAT & FORA & UnSplit & PCAT & FORA\\
    \hline
    layer 1 & 0.171 & 0.853 & \textbf{0.926} & 11.03 & 22.10 & \textbf{25.87} & 0.677 & 0.219 & \textbf{0.120}\\
    layer 2 & 0.101 & 0.642 & \textbf{0.830} & 10.48 & 17.29 & \textbf{22.19} & 0.689 & 0.432 & \textbf{0.252}\\
    layer 3 & 0.104 & 0.291 & \textbf{0.622} & 11.14 & 13.18 & \textbf{18.93} & 0.741 & 0.615 & \textbf{0.381}\\
    layer 4 & 0.108 & \textbf{0.121} & 0.030 & 8.62 & \textbf{11.08} & 10.45 & 0.722 & 0.676 & \textbf{0.628}\\
    \hline
\end{tabular}
}
\label{tab:dra_cifar}
\end{table}

\begin{table}[htbp]
\centering
\caption{SSIM, PSNR and LPIPS of the reconstructed images on CelebA among three attacks.}
\resizebox{\linewidth}{!}{
\begin{tabular}{c |c c c| ccc | ccc }
    \hline
    \multirow{2}{*}{\makecell{Split \\Point}} & \multicolumn{3}{c|}{SSIM$\uparrow$} & \multicolumn{3}{c|}{PSNR$\uparrow$} & \multicolumn{3}{c}{LPIPS$\downarrow$}\\
    & UnSplit & PCAT & FORA & UnSplit & PCAT & FORA & UnSplit & PCAT & FORA\\
    \hline
    layer 1 & 0.137 & 0.333 & \textbf{0.485} & 9.26 & 13.45 & \textbf{17.72} & 0.804 & 0.634 & \textbf{0.320}\\
    layer 2 & 0.170 & 0.316 & \textbf{0.476} & 8.70 & 12.05 & \textbf{17.11} & 0.747 & 0.653 & \textbf{0.381}\\
    layer 3 & 0.156 & 0.164 & \textbf{0.191} & 10.66 & 11.63 & \textbf{14.19} & 0.793 & 0.731 & \textbf{0.509}\\
    layer 4 & 0.084 & 0.092 & \textbf{0.192} & 7.94 & 10.60 & \textbf{13.00} & 0.804 & 0.738 & \textbf{0.621}\\
    \hline
\end{tabular}
}
\label{tab:dra_celeba}
\end{table}

\textbf{Feature Similarity.}
As shown in \cref{tab:feat_dis}, we measure the feature distance between the proxy clients built by UnSplit, PCAT, and FORA and the target client at layer 2. The results show that the substitute clients trained by our method exhibit more similar representation preferences to the target client. 
The basic optimization approach of UnSplit makes it difficult to regularize the feature space of the proxy client.
As for PCAT, it simply makes the smashed data generated by the pseudo model more favorable to the server model but fails to mimic the behavior of the client model. In contrast, FORA can impose stronger constraints in the feature space, which directly contributes to successful reconstruction.

% previous version
% Shown as \cref{tab:feat_dis}, we measure the feature distance between the clients trained by PCAT and our method and the target client in split 2. Although we share a similar attack framework with PCAT, thanks to our new perspective, we can impose stronger constraints in the feature space. This enables our trained substitute client to exhibit behavior more similar to the target. In fact, PCAT just makes the smashed data generated by the pseudo model more favorable for the server model but fails to imitate the client model’s representation preference.

\begin{table}[tbp]
\centering
\caption{Feature similarity measured by Mean Square Error and Cosine Similarity on CIFAR-10 and CelebA at layer 2.}
\resizebox{\linewidth}{!}{
\begin{tabular}{c |cc c| ccc}
    \hline
    \multirow{2}{*}{Method} & \multicolumn{3}{c|}{CIFAR-10} & \multicolumn{3}{c}{CelebA} \\
    & UnSplit & PCAT & FORA & UnSplit & PCAT & FORA\\
    \hline
    Mean Square Error$\downarrow$ & 1.041 & 0.528 & \textbf{0.274} & 50.773 & 1.353 & \textbf{0.753}\\
    Cosine Similarity$\uparrow$ & 0.200 & 0.592 & \textbf{0.810} & 0.333 & 0.480 & \textbf{0.778}\\
    \hline
\end{tabular}
}
\label{tab:feat_dis}
\end{table}

\subsection{Effect of Auxiliary Dataset}
\label{subsec:Effect of Auxiliary Dataset}
Next, we analyze the effect of several important factors regarding the auxiliary dataset on attack performance. We first explore the impact of the fitting level of substitute models by varying the size of the auxiliary data. Then, we discuss the impact of the presence of a more significant distribution shift, i.e., the absence of some categories, between the auxiliary and target samples. Finally, we relax the major assumption about the adversary, namely that the server has access to the similarly distributed auxiliary dataset. We set the split point at layer 2 for ablation, and the full experimental results are provided in \cref{appendix_additional_results_data}.

% previous version
% In this section, we evaluated the impact of the auxiliary dataset on our method. Now, let's consider some possible scenarios for server: 1) the server cannot collect enough auxiliary data, 2) there is a more significant distribution shift between the auxiliary data and the target dataset, such as class-missing.  we evaluated the impact of those scenarios on our experiments. Furthermore, we evaluated the impact of our results when relaxing our attack assumption, i.e., when the server can obtain an i.i.d. auxiliary dataset. In this section, we set cut point at split2.

\begin{figure}[t]
    \centering
    \hfill
    \subfloat[CIFAR-10]{
        \begin{overpic}[width=1.38in]{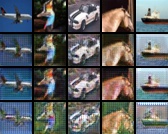}
            \put(-19,67){\footnotesize Truth}
            \put(-18,45){\footnotesize 5000}
            \put(-18,25){\footnotesize 2500}
            \put(-18,5){\footnotesize 1000}
        \end{overpic}
    }
    \hfill
    \hfill
    \subfloat[CelebA]{
        \begin{overpic}[width=1.38in]{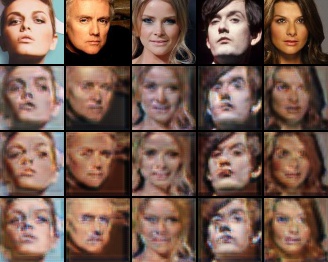}
            \put(-19,67){\footnotesize Truth}
            \put(-21,45){\footnotesize 10000}
            \put(-18,25){\footnotesize 5000}
            \put(-18,5){\footnotesize 2000}
        \end{overpic}
    }
\caption{Effects of  varying auxiliary data size on FORA
 performed on CIFAR-10 and CelebA at layer 2.}
\label{fig:dataset_quantity_pics}
\end{figure}
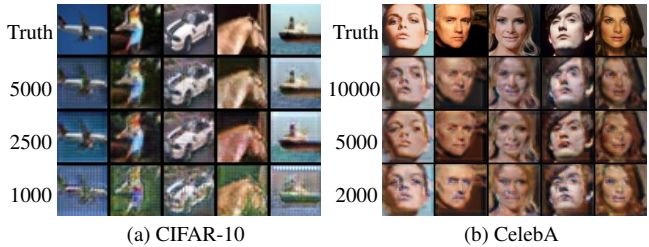

\textbf{Auxiliary Set Size.}
As shown in \cref{fig:dataset_quantity_pics}, when we reduce the size of the auxiliary dataset to half of the previous one, the attack performance of FORA remains almost unchanged. When we further reduce the number of auxiliary samples to 20\%, the quality of the reconstructed images decreases slightly but still preserves the full outline and most of the details. In that case, the percentage of the public auxiliary dataset is very small compared to the huge private training set (50,000 for CIFAR-10 and 162770 for CelebA), only 2\% and 1.2\%, respectively. This implies that even with a rather limited auxiliary dataset, FORA is still able to effectively reconstruct the client's training samples.

% previous version
% We explore the impact of the auxiliary dataset on FORA's results at split2. As shown in  \cref{tab:dataset_quantity}, when we reduce the number of auxiliary samples to half of the previous amount, the reconstructed images by FORA remain nearly constant. As we further decrease the number of auxiliary samples to 1000/2000, the quality of the reconstructed images shows a slight decline, but still retains complete outlines and most details. Compared to the vast privacy data (5000, 162770), our auxiliary dataset accounts for a very small proportion, only 2\% and 1.2\%, respectively. This implies that even with a limited auxiliary dataset, the server can still effectively reconstruct the client's training data. More reconstructed images is presented in Appendix \cref{fig:additional_cifar_quantity} and \cref{fig:additional_celeba_quantity}.

% \begin{table}[ht]
% \centering
% \caption{Effect of the Quantity of Auxiliary Dataset.}
% \resizebox{\linewidth}{!}{
% \begin{tabular}{c | ccc| ccc}
%     \hline
%     \multirow{2}{*}{\makecell{Dataset \\Size}} & \multicolumn{3}{c|}{CIFAR-10} & \multicolumn{3}{c}{CelebA}\\
%     & 5000 & 2500 & 1000 & 10000 & 5000 & 2000\\
%     \hline
%     SSIM$\uparrow$ & \textbf{0.830} & 0.817 & 0.612 &\textbf{0.476} & 0.452 & 0.465\\
%     PSNR$\uparrow$ & \textbf{22.19} & 22.14 & 16.69 & 17.11 & 16.71 & \textbf{17.18}\\
%     LPIPS$\downarrow$ & \textbf{0.252} & 0.259 & 0.483 & \textbf{0.381} & 0.397 & 0.416\\
%     \hline
% \end{tabular}
% }
% \label{tab:dataset_quantity}
% \end{table}

\begin{table}[h]
\centering
\caption{Effect of absence of categories on FORA
 performed on CIFAR-10  at layer 2.}
\resizebox{\linewidth}{!}{
\begin{tabular}{c| ccc c@{\hspace{0pt}} c@{\hspace{0pt}} c@{\hspace{0pt}} c@{\hspace{0pt}} c@{\hspace{0pt}}}
    \hline
    \makecell{Absent \\ Categories} & SSIM$\uparrow$ & PSNR$\uparrow$ & LPIPS$\downarrow$
    & \begin{minipage}[c]{0.14\columnwidth}
    \centering
    {\includegraphics[width=1.0\textwidth]{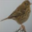}}
    \end{minipage}
    & \begin{minipage}[c]{0.14\columnwidth}
    \centering
    {\includegraphics[width=1.0\textwidth]{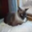}}
    \end{minipage}
    & \begin{minipage}[c]{0.14\columnwidth}
    \centering
    {\includegraphics[width=1.0\textwidth]{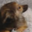}}
    \end{minipage}
    & \begin{minipage}[c]{0.14\columnwidth}
    \centering
    {\includegraphics[width=1.0\textwidth]{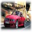}}
    \end{minipage}
    & \begin{minipage}[c]{0.14\columnwidth}
    \centering
    {\includegraphics[width=1.0\textwidth]{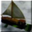}}
    \end{minipage}
    \\
    \hline
    Living & 0.768 & 20.44 & 0.300
    & \begin{minipage}[c]{0.14\columnwidth}
    \centering
    {\includegraphics[width=1.0\textwidth]{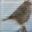}}
    \end{minipage}
    & \begin{minipage}[c]{0.14\columnwidth}
    \centering
    {\includegraphics[width=1.0\textwidth]{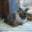}}
    \end{minipage}
    & \begin{minipage}[c]{0.14\columnwidth}
    \centering
    {\includegraphics[width=1.0\textwidth]{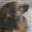}}
    \end{minipage}
    & \begin{minipage}[c]{0.14\columnwidth}
    \centering
    {\includegraphics[width=1.0\textwidth]{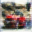}}
    \end{minipage}
    & \begin{minipage}[c]{0.14\columnwidth}
    \centering
    {\includegraphics[width=1.0\textwidth]{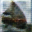}}
    \end{minipage}
    \\
    Non-living & 0.732 & 18.43 & 0.395
    & \begin{minipage}[c]{0.14\columnwidth}
    \centering
    {\includegraphics[width=1.0\textwidth]{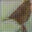}}
    \end{minipage}
    & \begin{minipage}[c]{0.14\columnwidth}
    \centering
    {\includegraphics[width=1.0\textwidth]{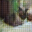}}
    \end{minipage}
    & \begin{minipage}[c]{0.14\columnwidth}
    \centering
    {\includegraphics[width=1.0\textwidth]{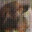}}
    \end{minipage}
    & \begin{minipage}[c]{0.14\columnwidth}
    \centering
    {\includegraphics[width=1.0\textwidth]{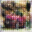}}
    \end{minipage}
    & \begin{minipage}[c]{0.14\columnwidth}
    \centering
    {\includegraphics[width=1.0\textwidth]{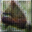}}
    \end{minipage}
    \\
    \hline
\end{tabular}
}
\label{tab:class_missing}
\end{table}

\textbf{Absence of Categories.}
It is likely that the adversary's public auxiliary data misses some semantic classes of the private data distribution. To model this situation, we create two special auxiliary datasets for CIFAR-10, one containing ``Living'' items (birds, cats, etc.), and the other containing ``Non-living'' items (airplanes, cars, etc.), both with 5,000 randomly sampled samples from CINIC-10. As presented in \cref{tab:class_missing}, even if a class is absent from the auxiliary dataset, FORA can still reconstruct samples of that class. In fact, FORA focuses on stealing the mapping relationship between client inputs and smashed data and therefore does not require class alignment. We observe that the absence of the ``Non-living'' category leads to a moderate degradation in the reconstruction results. We believe that the reason behind this phenomenon is that the greater variation of classes within the ``Non-living'' category helps to increase the generalization level of the substitute client, which in turn facilitates improved attack performance.

% previous version
% We discuss the impact of class-missing in the auxiliary dataset on FORA. For CIFAR-10, we set up two auxiliary datasets, one containing only living items (bird,cat,dog,deer,frog,horse) and the other containing non-living items (airplane,automobile,ship,truck), randomly sampled from CINIC-10. We maintain the data size of the both sets at 5000. As shown in \cref{tab:class_missing}, even when a certain class is missing in the auxiliary dataset, FORA can still reconstruct data from that class. In fact, FORA just steals the mapping relationship between client inputs and smashed data, therefore  class alignment is not necessary. We observe that missing the ``non-living'' class leads to a more significant drop in reconstruction results. We believe this is because the class variation within "non-living" categories is larger, which helps improve the generalization of the substitute client. More reconstruction results shown in  Appendix \cref{fig:additional_class_missing}.

\textbf{Distribution Shift.}
Here we further analyze the impact of the auxiliary dataset distribution on FORA. 
In contrast to our default experimental setup, we selected 5000 and 10000 images from the original testing sets of CIFAR-10 and CelebA, respectively, as the auxiliary datasets with the same distribution. 
As shown in \cref{tab:iid}, a more similar distribution can facilitate substitute clients stealing the representation preference, resulting in better reconstruction performance.
We observe that the attack results on the facial dataset are more vulnerable to the data distribution shift compared to the object dataset.
One possible reason is that tasks related to facial datasets are more sensitive to variations in sampling methods and alignment conditions across different datasets.
For object datasets, due to substantial distribution variation between different categories of themselves, \eg ranging from animals to vehicles, which contributes to their robustness in handling distribution shifts.

\begin{table}[t]
\centering
\caption{Effects of auxiliary dataset distribution shift on FORA performed on CIFAR-10 and CelebA at layer 2. ``Different'' represents auxiliary data sampled from CINIC-10, and FFHQ respectively, and  ``Same'' means auxiliary dataset come from their original test set.}
\resizebox{0.78\linewidth}{!}{
\begin{tabular}{c | cc| cc}
    \hline
    \multirow{2}{*}{\makecell{Dataset \\Size}} & \multicolumn{2}{c|}{CIFAR-10} & \multicolumn{2}{c}{CelebA}\\
    & Different & Same & Different & Same\\
    \hline
    SSIM$\uparrow$ & 0.830 & 0.832 & 0.476 & 0.777\\
    PSNR$\uparrow$ & 22.19 & 22.78 & 17.11 & 21.55\\
    LPIPS$\downarrow$ & 0.252 & 0.207 & 0.381 & 0.264\\
    \hline
\end{tabular}
}
\label{tab:iid}
\end{table}

\subsection{Effect of Substitute Client Structure}
\label{subsec:Effect of Substitute Client Structure}

\begin{figure}[b]
\centering
    \subfloat[SSIM$\uparrow$]{\includegraphics[width=1.58in]{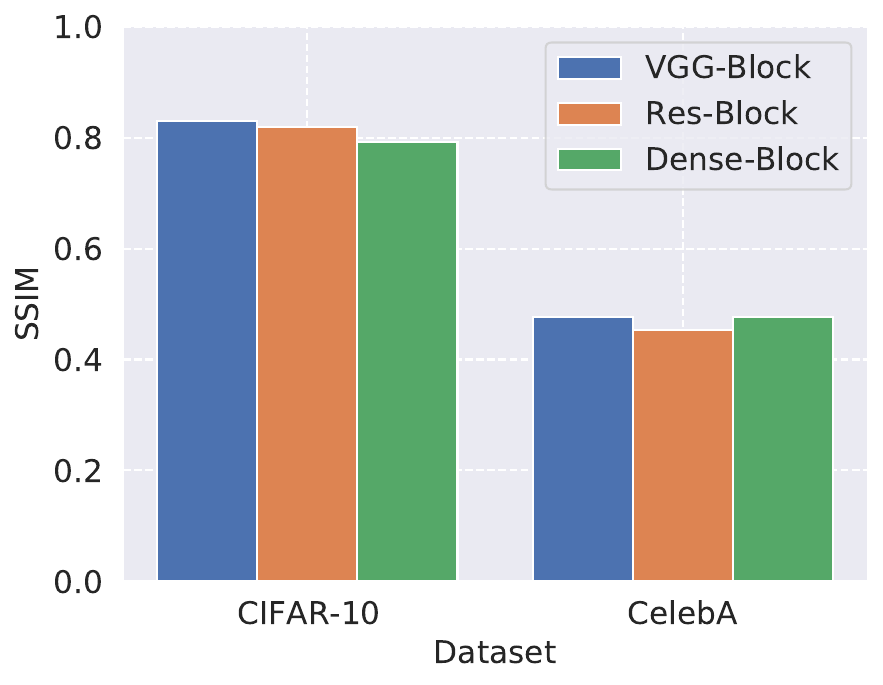}}
    % \subfloat[PSNR]{\includegraphics[width=1.5in]{img/evaluation/structure/psnr_bar.pdf}}
    \subfloat[LPIPS$\downarrow$]{\includegraphics[width=1.58in]{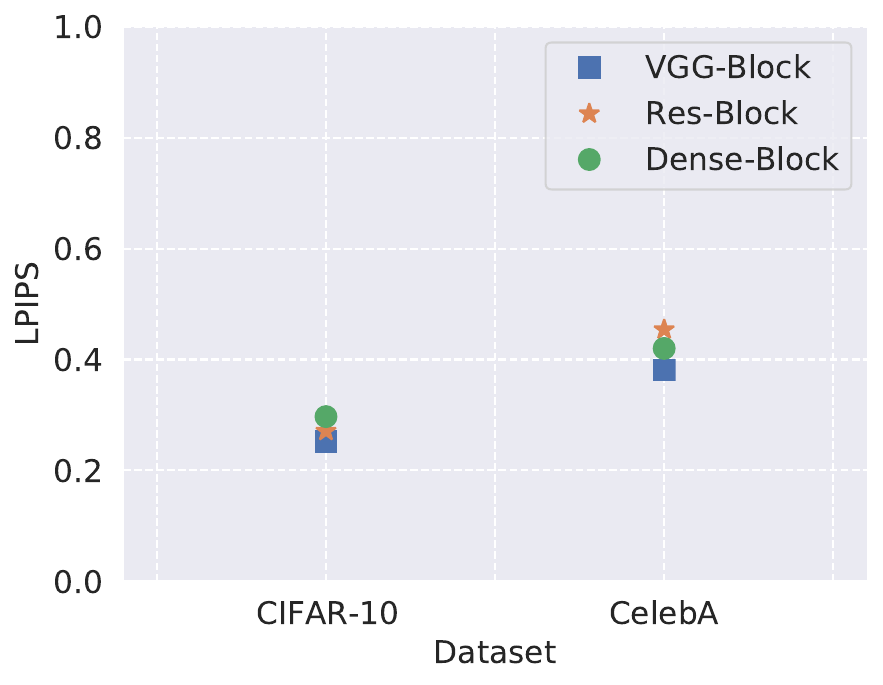}}
\caption{Effect for FORA with varying substitute model architectures on both datasets at layer 2.}
\label{fig:ssim_structure}
\end{figure}

After validating the impact of the auxiliary dataset, here we are interested in the impact of substitute client architectures on FORA. We chose three different model structures as attack variants: the VGG block \cite{vgg}, the ResNet block \cite{resnet}, and the DenseNet block \cite{densenet}. As can be seen in \cref{fig:ssim_structure}, the SSIM and LPIPS quantization results for the reconstructed images remain similar. This indicates that the extracted representation preferences on the basis of MK-MDD and Discriminator are close to that of the target client, despite the fact that the substitute clients use different architectures. Additional results are shown in \cref{appendix_result_arc}.

% previous version
% Our method proves that server does not require knowledge of the client's  architecture. However, since server has no any prior knowledge about the client, it may adopt a substitute client with any architecture.
% Then we further investigate the impact of different substitute client architectures for FOSA. We choose three popular and common network architectures: vgg block, resnet block, and desnet block as the  baselines.
% \textbf{Experiment Results.} We find that the ssim and lpips of reconstructed images shown in \cref{fig:ssim_structure} are almost invariant on different architectures, which proves that FORA truly does not depend on prior knowledge of the target client architecture. We believe this is because current neural networks have sufficient parameter capacity and benign  design, so they have enough ability to imitate client behavior.

% %--------------------------------------------------------------
% \begin{figure}[t]
% \centering
%     \subfloat[SSIM]{\includegraphics[width=1.62in]{img/evaluation/structure/ssim_bar.pdf}}
%     % \subfloat[PSNR]{\includegraphics[width=1.5in]{img/evaluation/structure/psnr_bar.pdf}}
%     \subfloat[LPIPS]{\includegraphics[width=1.62in]{img/evaluation/structure/lpips_line-new.pdf}}
% \caption{Effect for FORA with varying substitute model architectures on both datasets at layer 2.}
% \label{fig:ssim_structure}
% \end{figure}
% %--------------------------------------------------------------

\subsection{Counter Defense Techniques}
\label{subsec:Counter Defense Techniques}

% In SL, smashed data from clients may contain rich spatial information. Therefore, 
There have been a number of defenses aimed at perturbing the smashed data claiming that they can reduce the risk of privacy leakage in SL to a certain extent. We select three well-known defense techniques, i.e., distance correlation minimization \cite{vepakomma2020nopeek,szekely2007measuring,vepakomma2019reducing}, differential privacy \cite{abadi2016deep}, and noise obfuscation \cite{titcombe2021practical}, to evaluate the effectiveness of FORA. \cref{tab:defense_cifar} shows the limited impact of these defenses on FORA. See \cref{appendix_defense_details} for more details on defense techniques. See \cref{appendix_defense_celeba} for more defense results and discussions about possible adaptive defenses.

% previous version
% To mitigate various inference attacks on deep learning models, several defense approaches have been proposed \cite{vepakomma2019reducing,szekely2007measuring,vepakomma2020nopeek,abadi2016deep,titcombe2021practical}. 
% In SL, the smashed data from the client contains abundant information. Thus, perturbing the smashed data can help reduce the risk of privacy leakage to some extent. For evaluation, we have selected three defensive techniques that can impact the smashed data: noppek \cite{vepakomma2020nopeek,szekely2007measuring,vepakomma2019reducing}, differential privacy \cite{abadi2016deep} and additive noise \cite{titcombe2021practical}. More details about defenses are provided in Appendix A?????.

\textbf{Distance Correlation Minimization (DCOR).} DCOR can uncorrelate irrelevant and sensitive features from the smashed data associated with the target client, which results in a lack of detailed expression of the input data in the representation preferences learned by the substitute client, especially in colors. However, FORA retains the ability to reconstruct the structural details of the private image.

% previous version
% \textbf{Nopeek} can decorrelate irrelevant sensitive information from the related smashed data, which results in the substitute client as well as the target client sharing similar representation preferences, lacking a detailed representation of the input data, especially in terms of color.
% Although  Nopeek will introduce confusion into the color information of reconstructed images, FORA retains the ability to reconstruct outlines and a substantial portion of the details.

\textbf{Differential Privacy (DP).} DP protects training data privacy by adding carefully crafted Laplace noise to the gradients. However, the effectiveness of DP against FORA is very limited under all privacy budgets. When the test accuracy of the model is reduced by nearly 10\% (the functionality is severely damaged), the SSIM of the reconstructed samples still reaches about 75\% of the original. This trade-off between classification accuracy and defense strength makes DP not feasible for practical applications of SL.

% \textbf{Differential Privacy (DP).} DP protects training data privacy by adding carefully crafted Laplace noise to the gradients. However, the effectiveness of DP against FORA is very limited under all privacy budgets. That is, when the test accuracy of the SL model is reduced by nearly 10\% (the classification functionality is severely damaged), the SSIM of the reconstructed samples still reaches about 75\% of the original. This trade-off between classification accuracy and defense strength makes DP not feasible for practical applications of SL.

% previous version
% \textbf{Differential Privacy (DP)} protects model privacy by adding carefully crafted noise to gradients, but this approach can significantly compromise the effectiveness of SL. However, the defense mechanism against FORA is highly limited. Even when the model accuracy decreases by nearly 10\%, the SSIM of the reconstructed images only decreases by 0.202.

\textbf{Noise Obfuscation (NO).} NO is a direct defense to destroy the mapping relationship between smashed and input data. We observe that on the one hand, the noise of a small scale enhances the generalization level of the SL model to maintain or even improve the classification accuracy, on the other hand raising the noise scale helps to introduce deviations to the features extracted from the target client, making it more difficult to learn the representations and reconstruct the data for FORA.

% previous version
% \textbf{Noise Obfuscation} is a direct defense to disrupt the mapping relationship between smashed data and input data, posing a significant challenge for DRA. 
% However, FORA still achieves satisfactory reconstruction results, even when the noise scale is set to 5, the reconstructed images remain distinguishable.
% We observed a fluctuation in testing accuracy with different defense hyperparameters, showing an initial increase followed by a decrease. We attribute this to moderate noise enhancing the model's generalization, but when the noise scale becomes too large, it starts to adversely affect the model's normal training process.

\begin{table}[t]
\centering
\caption{Effect of utility and FORA performance against three defense techniques on CIFAR-10 at layer 2.}
\resizebox{\linewidth}{!}{
\begin{tabular}{c c cccc@{\hspace{0pt}} c@{\hspace{0pt}} c@{\hspace{0pt}} c@{\hspace{0pt}}}
    \hline
    \makecell{Defense \\ Hyperparam} & Test Acc (\%) & SSIM$\uparrow$ & PSNR$\uparrow$ & LPIPS$\downarrow$
    & \begin{minipage}[c]{0.14\columnwidth}
    \centering
    {\includegraphics[width=1.0\textwidth]{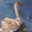}}
    \end{minipage}
    & \begin{minipage}[c]{0.14\columnwidth}
    \centering
    {\includegraphics[width=1.0\textwidth]{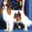}}
    \end{minipage}
    & \begin{minipage}[c]{0.14\columnwidth}
    \centering
    {\includegraphics[width=1.0\textwidth]{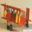}}
    \end{minipage}
    & \begin{minipage}[c]{0.14\columnwidth}
    \centering
    {\includegraphics[width=1.0\textwidth]{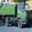}}
    \end{minipage}
    \\
    \hline
    \makecell{0 \\(w/o defense)} & 71.25 & 0.830 & 22.19 & 0.252
    & \begin{minipage}[c]{0.14\columnwidth}
    \centering
    {\includegraphics[width=1.0\textwidth]{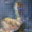}}
    \end{minipage}
    & \begin{minipage}[c]{0.14\columnwidth}
    \centering
    {\includegraphics[width=1.0\textwidth]{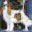}}
    \end{minipage}
    & \begin{minipage}[c]{0.14\columnwidth}
    \centering
    {\includegraphics[width=1.0\textwidth]{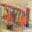}}
    \end{minipage}
    & \begin{minipage}[c]{0.14\columnwidth}
    \centering
    {\includegraphics[width=1.0\textwidth]{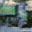}}
    \end{minipage}
    \\
    \hline
    \multicolumn{5}{c}{\textbf{DCOR ($\alpha$)}} & & & &
    \\ 
    \hline
    0.2 & 70.91 & 0.692 & 17.91 & 0.360
    & \begin{minipage}[c]{0.14\columnwidth}
    \centering
    {\includegraphics[width=1.0\textwidth]{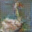}}
    \end{minipage}
    & \begin{minipage}[c]{0.14\columnwidth}
    \centering
    {\includegraphics[width=1.0\textwidth]{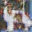}}
    \end{minipage}
    & \begin{minipage}[c]{0.14\columnwidth}
    \centering
    {\includegraphics[width=1.0\textwidth]{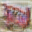}}
    \end{minipage}
    & \begin{minipage}[c]{0.14\columnwidth}
    \centering
    {\includegraphics[width=1.0\textwidth]{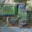}}
    \end{minipage}
    \\
    0.5 & 70.06 & 0.628 & 15.99 & 0.441
    & \begin{minipage}[c]{0.14\columnwidth}
    \centering
    {\includegraphics[width=1.0\textwidth]{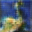}}
    \end{minipage}
    & \begin{minipage}[c]{0.14\columnwidth}
    \centering
    {\includegraphics[width=1.0\textwidth]{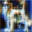}}
    \end{minipage}
    & \begin{minipage}[c]{0.14\columnwidth}
    \centering
    {\includegraphics[width=1.0\textwidth]{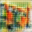}}
    \end{minipage}
    & \begin{minipage}[c]{0.14\columnwidth}
    \centering
    {\includegraphics[width=1.0\textwidth]{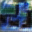}}
    \end{minipage}
    \\
    0.8 & 69.72 & 0.563 & 15.40 & 0.471
    & \begin{minipage}[c]{0.14\columnwidth}
    \centering
    {\includegraphics[width=1.0\textwidth]{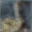}}
    \end{minipage}
    & \begin{minipage}[c]{0.14\columnwidth}
    \centering
    {\includegraphics[width=1.0\textwidth]{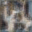}}
    \end{minipage}
    & \begin{minipage}[c]{0.14\columnwidth}
    \centering
    {\includegraphics[width=1.0\textwidth]{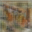}}
    \end{minipage}
    & \begin{minipage}[c]{0.14\columnwidth}
    \centering
    {\includegraphics[width=1.0\textwidth]{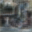}}
    \end{minipage}
    \\
    \hline
    \multicolumn{5}{c}{\textbf{DP ($\epsilon$)}} & & & &
    \\
    \hline
    $+\infty$ & 69.68 & 0.823 & 22.36 & 0.225
    & \begin{minipage}[c]{0.14\columnwidth}
    \centering
    {\includegraphics[width=1.0\textwidth]{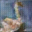}}
    \end{minipage}
    & \begin{minipage}[c]{0.14\columnwidth}
    \centering
    {\includegraphics[width=1.0\textwidth]{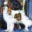}}
    \end{minipage}
    & \begin{minipage}[c]{0.14\columnwidth}
    \centering
    {\includegraphics[width=1.0\textwidth]{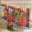}}
    \end{minipage}
    & \begin{minipage}[c]{0.14\columnwidth}
    \centering
    {\includegraphics[width=1.0\textwidth]{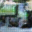}}
    \end{minipage}
    \\
    100  & 63.05 & 0.711 & 20.36 & 0.394
    & \begin{minipage}[c]{0.14\columnwidth}
    \centering
    {\includegraphics[width=1.0\textwidth]{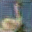}}
    \end{minipage}
    & \begin{minipage}[c]{0.14\columnwidth}
    \centering
    {\includegraphics[width=1.0\textwidth]{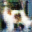}}
    \end{minipage}
    & \begin{minipage}[c]{0.14\columnwidth}
    \centering
    {\includegraphics[width=1.0\textwidth]{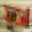}}
    \end{minipage}
    & \begin{minipage}[c]{0.14\columnwidth}
    \centering
    {\includegraphics[width=1.0\textwidth]{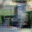}}
    \end{minipage}
    \\
    10  & 61.93 & 0.621 & 18.03 & 0.487
    & \begin{minipage}[c]{0.14\columnwidth}
    \centering
    {\includegraphics[width=1.0\textwidth]{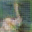}}
    \end{minipage}
    & \begin{minipage}[c]{0.14\columnwidth}
    \centering
    {\includegraphics[width=1.0\textwidth]{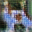}}
    \end{minipage}
    & \begin{minipage}[c]{0.14\columnwidth}
    \centering
    {\includegraphics[width=1.0\textwidth]{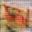}}
    \end{minipage}
    & \begin{minipage}[c]{0.14\columnwidth}
    \centering
    {\includegraphics[width=1.0\textwidth]{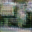}}
    \end{minipage}
    \\
    % 0.6 & 95.54 & 61.71 & 0.670 & 19.16 & 0.463
    % & \begin{minipage}[c]{0.14\columnwidth}
    % \centering
    % {\includegraphics[width=1.0\textwidth]{img/evaluation/defense/bird_dp_2.0.png}}
    % \end{minipage}
    % & \begin{minipage}[c]{0.14\columnwidth}
    % \centering
    % {\includegraphics[width=1.0\textwidth]{img/evaluation/defense/dog_dp_2.0.png}}
    % \end{minipage}
    % & \begin{minipage}[c]{0.14\columnwidth}
    % \centering
    % {\includegraphics[width=1.0\textwidth]{img/evaluation/defense/plane_dp_2.0.png}}
    % \end{minipage}
    % & \begin{minipage}[c]{0.14\columnwidth}
    % \centering
    % {\includegraphics[width=1.0\textwidth]{img/evaluation/defense/truck_dp_2.0.png}}
    % \end{minipage}
    % \\
    \hline
    \multicolumn{5}{c}{\textbf{NO ($\sigma$)}} & & & &
    \\
    \hline
    1.0 & 74.39 & 0.640 & 17.29 & 0.367
    & \begin{minipage}[c]{0.14\columnwidth}
    \centering
    {\includegraphics[width=1.0\textwidth]{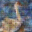}}
    \end{minipage}
    & \begin{minipage}[c]{0.14\columnwidth}
    \centering
    {\includegraphics[width=1.0\textwidth]{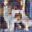}}
    \end{minipage}
    & \begin{minipage}[c]{0.14\columnwidth}
    \centering
    {\includegraphics[width=1.0\textwidth]{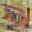}}
    \end{minipage}
    & \begin{minipage}[c]{0.14\columnwidth}
    \centering
    {\includegraphics[width=1.0\textwidth]{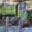}}
    \end{minipage}
    \\
    2.0 & 73.14 & 0.583 & 16.29 & 0.444
    & \begin{minipage}[c]{0.14\columnwidth}
    \centering
    {\includegraphics[width=1.0\textwidth]{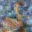}}
    \end{minipage}
    & \begin{minipage}[c]{0.14\columnwidth}
    \centering
    {\includegraphics[width=1.0\textwidth]{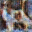}}
    \end{minipage}
    & \begin{minipage}[c]{0.14\columnwidth}
    \centering
    {\includegraphics[width=1.0\textwidth]{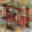}}
    \end{minipage}
    & \begin{minipage}[c]{0.14\columnwidth}
    \centering
    {\includegraphics[width=1.0\textwidth]{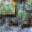}}
    \end{minipage}
    \\
    % 3.0 & 98.43 & 73.28 & 0.412 & 14.85 & 0.519
    % & \begin{minipage}[c]{0.14\columnwidth}
    % \centering
    % {\includegraphics[width=1.0\textwidth]{img/evaluation/defense/bird_noise_3.0.png}}
    % \end{minipage}
    % & \begin{minipage}[c]{0.14\columnwidth}
    % \centering
    % {\includegraphics[width=1.0\textwidth]{img/evaluation/defense/dog_noise_3.0.png}}
    % \end{minipage}
    % & \begin{minipage}[c]{0.14\columnwidth}
    % \centering
    % {\includegraphics[width=1.0\textwidth]{img/evaluation/defense/plane_noise_3.0.png}}
    % \end{minipage}
    % & \begin{minipage}[c]{0.14\columnwidth}
    % \centering
    % {\includegraphics[width=1.0\textwidth]{img/evaluation/defense/truck_noise_3.0.png}}
    % \end{minipage}
    % \\
    5.0 & 70.62 & 0.394 & 14.35 & 0.550
    & \begin{minipage}[c]{0.14\columnwidth}
    \centering
    {\includegraphics[width=1.0\textwidth]{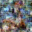}}
    \end{minipage}
    & \begin{minipage}[c]{0.14\columnwidth}
    \centering
    {\includegraphics[width=1.0\textwidth]{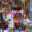}}
    \end{minipage}
    & \begin{minipage}[c]{0.14\columnwidth}
    \centering
    {\includegraphics[width=1.0\textwidth]{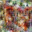}}
    \end{minipage}
    & \begin{minipage}[c]{0.14\columnwidth}
    \centering
    {\includegraphics[width=1.0\textwidth]{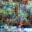}}
    \end{minipage}
    \\
    \hline
\end{tabular}
}
\label{tab:defense_cifar}
\end{table}
\section{Discussion and Conclusion}
\label{sec:Discussion and Conclusion}
% In this section, we first discuss the characteristics of our attack, and then we summarize this work.
%  our proposed method
In this section, we first discuss the potential improvement and scalability of FORA, then we summarize this work. We also show limitation and future work in \cref{more_discussions}.
% \textbf{FORA with Transfer Loss.} 
% If the server has access to the target task, in addition to the smashed data, it can further constrain the substitute client using the server model in each iteration. We define this function as transfer loss denoted by \begin{math}  \mathcal {L}_{trans} =  \mathcal{L}_{task} (F_S(f_s(X_{aux})),Y_{aux})  \end{math}, where  \begin{math}  \mathcal {L}_{task} \end{math} is the original training loss \eg cross entropy loss.
% The new reconstruction result is shown in...

\begin{figure}[t]
    \centering
    \hspace{22pt}
    \begin{overpic}[width=2.8in]{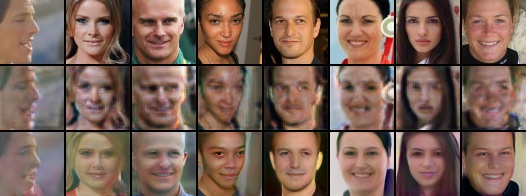}
        \put(-12.5,30){\footnotesize Truth}
        \put(-13,17){\footnotesize FORA}
        \put(-15,5){\footnotesize FORA-G}
    \end{overpic}
\caption{Reconstructed CelebA images of FORA and FORA-G, FOAR-G represents FORA combined with StyleGAN.}
\label{fig:gan}
\end{figure}

\textbf{Improvement using Generative Adversarial Networks.}
Li \etal \cite{glass} propose a novel StyleGAN-based reconstruction attack against split inference, and their research focus is orthogonal to our contribution. Therefore, the reconstruction task in FORA can be further optimized using pre-trained StyleGAN \cite{r13}. As shown in \cref{fig:gan}, the well-trained substitute client in FORA combined with StyleGAN optimization can provide additional improvements in reconstruction performance.

% previous version
% Li \etal \cite{glass}  introduced StyleGAN-based reconstruction attacks during the inference phase. We also demonstrate that a well trained substitute client can leverage StyleGAN to pose a greater privacy threat against SL.

\textbf{Attack on Label-Protected SL.}
Another popular setup for SL requires the client to keep the labels locally \cite{vepakomma2018split}, but this case does not have any influence on the implementation and performance of FORA. Since FORA is only related to the smashed data output from the target client, it does not depend on the server model as well as the training task.

% previous version
% We further analyze the effectiveness of FORA in another popular scenario of SL, where clients retain labels locally. In fact, this scenario has no impact on FORA, as our method is solely related to the smashed data output by the client and is independent of the server model.

\textbf{Conclusion.}
In this work, we propose a novel data reconstruction attack against SL, named Feature-Oriented Reconstruction Attack (FORA). Unlike all previous attack schemes, FORA enables a semi-honest server to secretly reconstruct the client's private training data with very little prior knowledge. Thanks to our new perspective of extracting representation preferences from smashed data, the server can contemporaneously train a substitute client that approximates the target client's behavior to conduct the attack. Our extensive experiments in various settings demonstrate the state-of-the-art performance of FORA. Due to its stealth and effectiveness, it poses a real privacy threat to SL. We hope our work can inspire future efforts to explore it in more practical SL, and we are eager to draw attention to more robust defense techniques.
% it poses a real privacy threat to SL. We We hope our work can inspires future efforts to explor  it in more practical SL scenarios, and we are eager to draw attention to more robust defense techniques.

% previous version
% In this work, we propose a novel privacy threat against SL, named FORA. Unlike traditional DRA against SL, FORA enables a semi-honest server to secretly reconstruct client training data with the minimum prior knowledge. Thanks to our new perspective on extracting feature preference from smashed data, the server can train a substitute client that closely mimics client behavior to perform the attack. Comprehensive experiments under various settings have demonstrated the effectiveness of FORA, which poses a real threat to SL  due to its stealthiness. I hope our work can inspire future exploration in more realistic scenarios, and we also aspire to draw attention to the privacy and security aspects SL.

% \balance 
{
    \small
    \bibliographystyle{ieeenat_fullname}
    \bibliography{ref}
}

\clearpage
\setcounter{page}{1}
\onecolumn
\maketitlesupplementary
\appendix

\section{Experimental Setup Details}
In this section, we provide more details about our experimental setup.

\subsection{Datasets}
\label{appendix_dataset}
We elaborate on the four datasets used in our main experiments and  \cref{tab:data splits} shows the details of datasets partitioning.

\textbf{CIFAR-10 \cite{cifar}.}
CIFAR-10 is a classification benchmark dataset comprising 60,000 3$\times$32$\times$32 images categorized into 10 classes. It features 50,000 training images and 10,000 testing images, evenly distributed among the classes.

\textbf{CINIC-10 \cite{cinic}.}
CINIC-10 extends CIFAR-10 by adding downsampled ImageNet samples in the same classes with CIFAR-10. 
Both datasets share the same classes, but CINIC-10 consists of 270,000 3$\times$32$\times$32 images.
In comparison to CIFAR-10, CINIC-10 presents a more complex and diverse distribution.

\textbf{CelebA \cite{celeba}.}
CelebA is a dataset related to facial attribute classification.
It includes 202,599 facial images from 10,177 different celebrities, and each image is associated with 40 different attribute labels.
In our experiment, we resize the images in the CelebA to 3$\times$64$\times$64.

\textbf{FFHQ \cite{ffhq}.}
FFHQ was originally designed as a benchmark for Generative Adversarial Networks which contains 70,000 facial images. The dataset exhibits rich diversity and noticeable variations in terms of age, ethnicity, and image backgrounds. 
As well as CelebA, we resize the images in FFHQ to 3$\times$64$\times$64.

%--------------------------------------------------------------
\begin{table}[htbp]
\centering
\caption{Details of the partitioning among different datasets.}
% \resizebox{\linewidth}{!}{
\begin{tabular}{|c |c c c| c | c |}
    \hline
    Target Model & Target Dataset & Train & Test & Auxiliary Dataset & Image Size\\
    \hline
    MobileNet & CIFAR-10 & 50000 & 10000 & CINIC-10 (5000)& 3$\times$32$\times$32\\
    ResNet-18 & CelebA   & 162770& 19962 & FFHQ (10000)& 3$\times$64$\times$64\\
    \hline
\end{tabular}
% }
\label{tab:data splits}
\end{table}
%--------------------------------------------------------------

\subsection{Model Architectures}
\label{appendix_arc}
\begin{figure}[ht]
\centering
\includegraphics[width=2.2in]{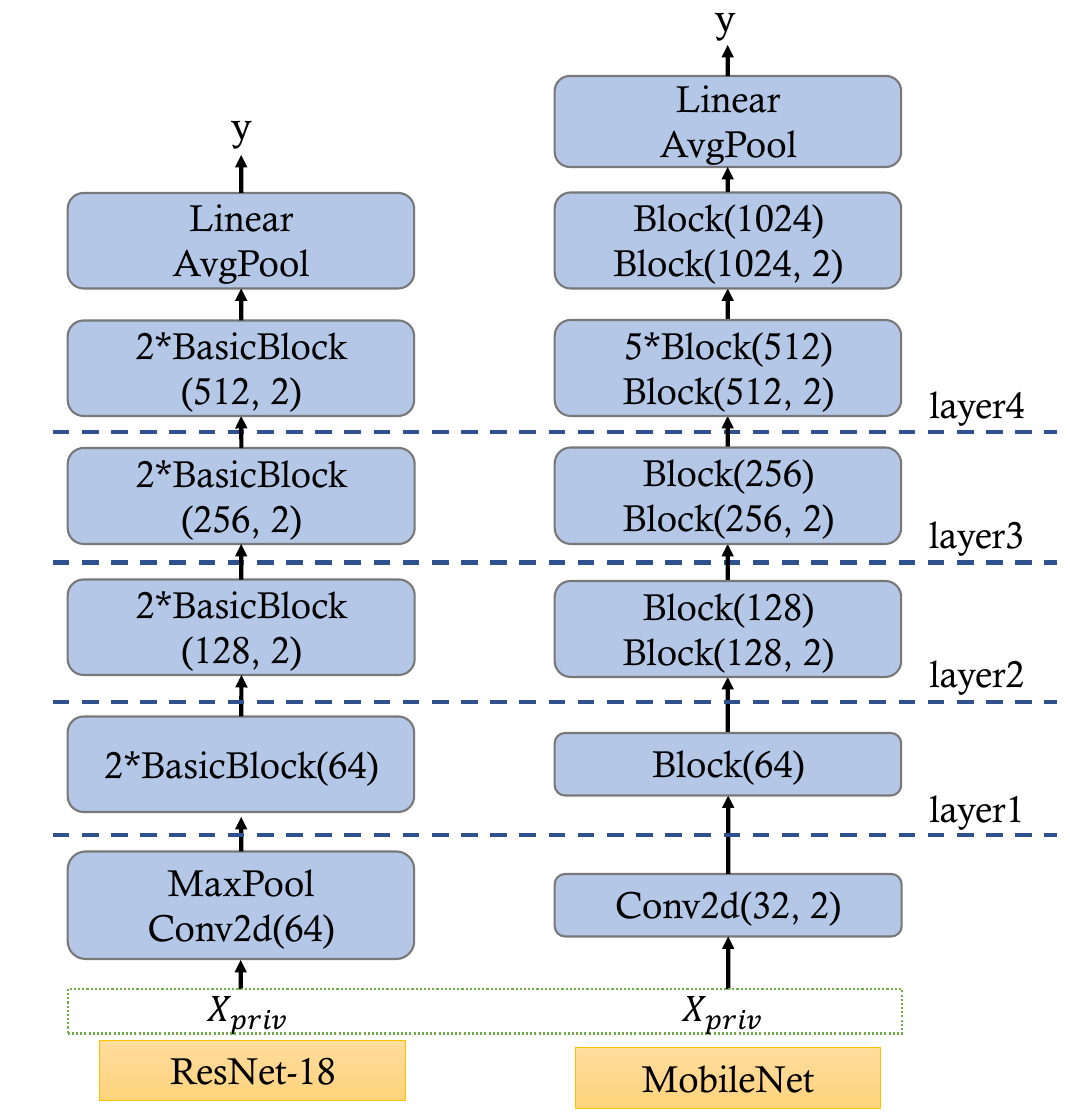}
\caption{Target models splitting settings in our experiments.}
\label{fig:targetmodel_partition}
\end{figure}

The detailed model architectures of the substitute client model, inverse network, and discriminator are shown in Table \ref{tab:model_details}. And \cref{fig:targetmodel_partition} illustrates the different split strategies towards the target model.

%--------------------------------------------------------------
\begin{table*}[t]
\label{other_arc}
\centering
\caption{Architectures of the substitute client, inverse network, and discriminator for two datasets in different splitting points.}
\resizebox{0.7\linewidth}{!}{
\begin{tabular}{c| ccc | ccc }
    \hline
    \multirow{2}{*}{Split Point} & \multicolumn{3}{c|}{CIFAR-10} & \multicolumn{3}{c}{CelebA}\\
    & $\hat{F_{c}}$ & $f^{-1}_{c}$ & $D$ & $\hat{F_{c}}$ & $f^{-1}_{c}$ & $D$\\
    \hline
    layer 1 
    & \makecell{2*Conv2d(32) \\[0.3ex] MaxPool}
    & \makecell{ConvTrans(256, 2) \\[0.3ex] Conv2d(256) \\[0.3ex] Tanh}
    & \makecell{Conv2d(128, 2) \\[0.3ex] Conv2d(256, 2) \\[0.3ex] 6*ResBlock(256) \\[0.3ex] Conv2d(256, 2) \\[0.3ex] Linear}
    & \makecell{2*Conv2d(64) \\[0.3ex] MaxPool}
    & \makecell{ConvTrans(256, 2) \\[0.3ex] Conv2d(256) \\[0.3ex] Tanh}
    & \makecell{Conv2d(128, 2) \\[0.3ex] 2*Conv2d(256, 2) \\[0.3ex] 6*ResBlock(256) \\[0.3ex] Conv2d(256, 2) \\[0.3ex] Linear}
    \\
    \hline
    layer 2 
    & \makecell{2*Conv2d(64) \\[0.3ex] MaxPool}
    & \makecell{ConvTrans(256, 2) \\[0.3ex] Conv2d(256) \\[0.3ex] Tanh}
    & \makecell{Conv2d(128, 2) \\[0.3ex] Conv2d(256, 2) \\[0.3ex] 6*ResBlock(256) \\[0.3ex] Conv2d(256, 2) \\[0.3ex] Linear}
    & \makecell{2*Conv2d(64) \\[0.3ex] MaxPool}
    & \makecell{ConvTrans(256, 2) \\[0.3ex] Conv2d(256) \\[0.3ex] Tanh}
    & \makecell{Conv2d(128, 2) \\[0.3ex] 2*Conv2d(256, 2) \\[0.3ex] 6*ResBlock(256) \\[0.3ex] Conv2d(256, 2) \\[0.3ex] Linear}
    \\
    \hline
    layer 3 
    & \makecell{2*Conv2d(64) \\[0.3ex] MaxPool \\[0.3ex] 2*Conv2d(128) \\[0.3ex] MaxPool}
    & \makecell{ConvTrans(256, 2) \\[0.3ex] ConvTrans(128, 2) \\[0.3ex] Conv2d(128) \\[0.3ex] Tanh}
    & \makecell{Conv2d(256, 2) \\[0.3ex] 6*ResBlock(256) \\[0.3ex] Conv2d(256, 2) \\[0.3ex] Linear}
    & \makecell{2*Conv2d(64) \\[0.3ex] MaxPool \\[0.3ex] 2*Conv2d(128) \\[0.3ex] MaxPool}
    & \makecell{ConvTrans(256, 2) \\[0.3ex] ConvTrans(128, 2) \\[0.3ex] Conv2d(128) \\[0.3ex] Tanh}
    & \makecell{2*Conv2d(256, 2) \\[0.3ex] 6*ResBlock(256) \\[0.3ex] Conv2d(256, 2) \\[0.3ex] Linear}
    \\
    \hline
    layer 4
    & \makecell{2*Conv2d(64) \\[0.3ex] MaxPool \\[0.3ex] 2*Conv2d(128) \\[0.3ex] MaxPool\\[0.3ex] 3*Conv2d(256) \\[0.3ex] MaxPool}
    & \makecell{ConvTrans(256, 2) \\[0.3ex] ConvTrans(128, 2) \\[0.3ex] ConvTrans(64, 2) \\[0.3ex] Conv2d(64) \\[0.3ex] Tanh}
    & \makecell{Conv2d(256) \\[0.3ex] 6*ResBlock(256) \\[0.3ex] Conv2d(256, 2) \\[0.3ex] Linear}
    & \makecell{2*Conv2d(64) \\[0.3ex] MaxPool \\[0.3ex] 2*Conv2d(128) \\[0.3ex] MaxPool\\[0.3ex] 3*Conv2d(256) \\[0.3ex] MaxPool}
    & \makecell{ConvTrans(256, 2) \\[0.3ex] ConvTrans(128, 2) \\[0.3ex] ConvTrans(64, 2) \\[0.3ex] Conv2d(64) \\[0.3ex] Tanh}
    & \makecell{Conv2d(256, 2) \\[0.3ex] Conv2d(256) \\[0.3ex] 6*ResBlock(256) \\[0.3ex] Conv2d(256, 2) \\[0.3ex] Linear}
    \\
    \hline
\end{tabular}
}
\label{tab:model_details}
\end{table*}
%--------------------------------------------------------------

% \clearpage
\section{Additional Data Reconstruction Results}
\label{appendix_additional_results}
We present more reconstruction results of FORA in this section.
\subsection{Comparison with Semi-Honest Attacks}
\label{appendix_additional_results_semi}
\begin{figure*}[hb]
    \centering
    \hfill
    \subfloat[UnSplit]{
        \begin{overpic}[width=2.0in]{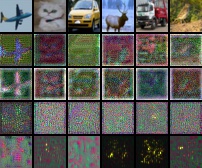}
            \put(-18,73){Truth}
            \put(-20,55){layer 1}
            \put(-20,39){layer 2}
            \put(-20,22){layer 3}
            \put(-20,6){layer 4}
        \end{overpic}
    }
    \hfill
    \subfloat[PCAT]{\includegraphics[width=2.0in]{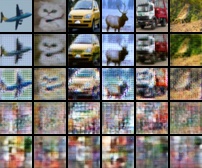}}
    \hfill
    \subfloat[FORA]{\includegraphics[width=2.0in]{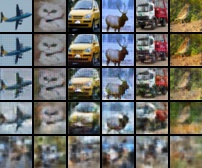}}
    \caption{Additional results of Unsplit, PCAT and FORA  on CIFAR-10 at all split points.}
    \label{fig:additional_cifar}
\end{figure*}

\begin{figure*}[htb]
    \centering
    \hfill
    \subfloat[UnSplit]{
        \begin{overpic}[width=2.0in]{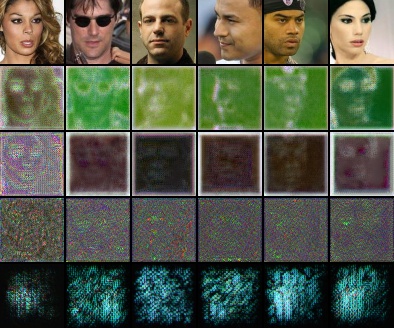}
            \put(-18,73){Truth}
            \put(-20,55){layer 1}
            \put(-20,39){layer 2}
            \put(-20,22){layer 3}
            \put(-20,6){layer 4}
        \end{overpic}
    }
    \hfill
    \subfloat[PCAT]{\includegraphics[width=2.0in]{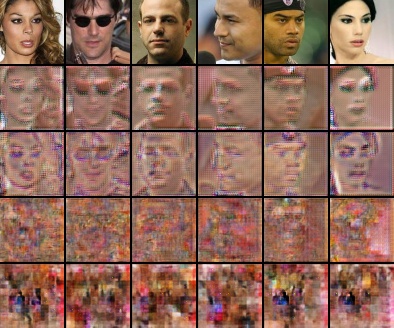}}
    \hfill
    \subfloat[FORA]{\includegraphics[width=2.0in]{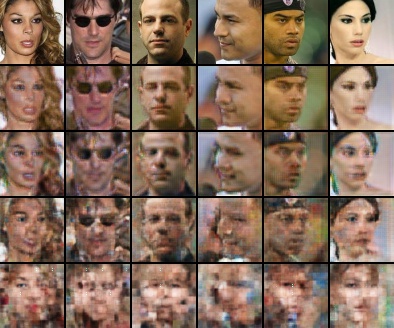}}
    \caption{Additional results of Unsplit, PCAT and FORA  on CelebA at all split points.}
    \label{fig:additional_celeba}
\end{figure*}

The full results of the comparison between UnSplit, PCAT, and FORA are presented in \cref{fig:additional_cifar} and \cref{fig:additional_celeba}.

%--------------
%------------

%--------------
%------------

\subsection{Effect of Auxiliary Dataset}
\label{appendix_additional_results_data}

\begin{figure}[ht]
\centering
% \hspace{16.2pt}
\begin{overpic}[width=2.8in]{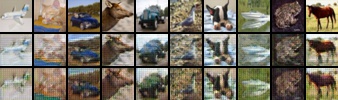}
    \put(-10,23){\footnotesize Truth}
    \put(-7,13){\footnotesize \ding{172}}
    \put(-7,4){\footnotesize \ding{173}}
\end{overpic}
\caption{Additional results of the absence of categories on CIFAR-10. Row \ding{172} means absence class of living. Row \ding{173} means absence class of non-living.}
\label{fig:additional_class_missing}
\end{figure}

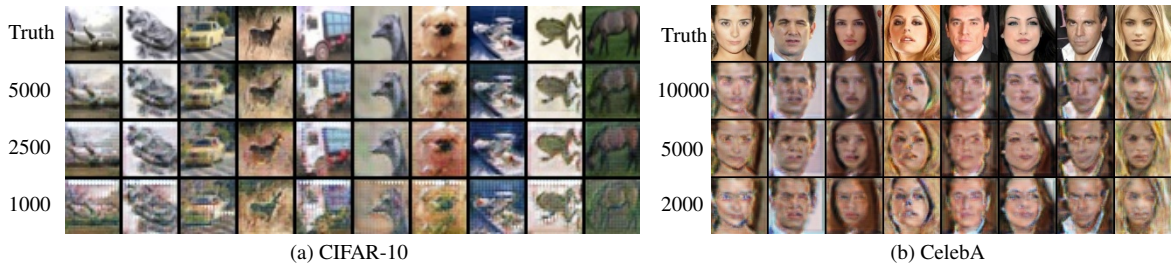
\begin{figure}[h]
\centering
% \hspace{1.5pt}
\subfloat[CIFAR-10]{
\begin{overpic}[width=3.0in]{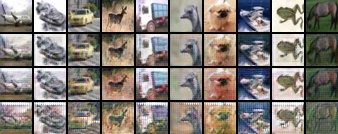}
    \put(-10,34){\footnotesize Truth}
    \put(-10,24){\footnotesize 5000}
    \put(-10,14){\footnotesize 2500}
    \put(-10,4){\footnotesize 1000}
\end{overpic}
}
\hspace{20pt}
\subfloat[CelebA]{
\begin{overpic}[width=2.4in]{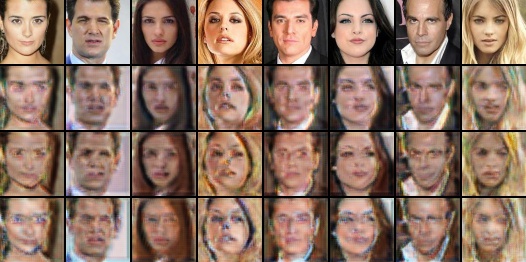}
    \put(-11,42){\footnotesize Truth}
    \put(-12,30){\footnotesize 10000}
    \put(-11,17){\footnotesize 5000}
    \put(-11,5){\footnotesize 2000}
\end{overpic}
}
\caption{Additional results of varying auxiliary data size on FORA performed on CIFAR-10 and CelebA.}
\label{fig:additional_quantity}
\end{figure}

\begin{figure}[ht]
\centering
% \hspace{20pt}
\subfloat[CIFAR-10]{
\begin{overpic}[width=3.0in]{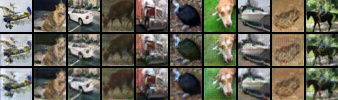}
    \put(-12,23){\footnotesize Truth}
    \put(-15.5,13){\footnotesize Different}
    \put(-11.5,4){\footnotesize Same}
\end{overpic}
}
\hspace{24pt}
\subfloat[CelebA]{
\begin{overpic}[width=2.4in]{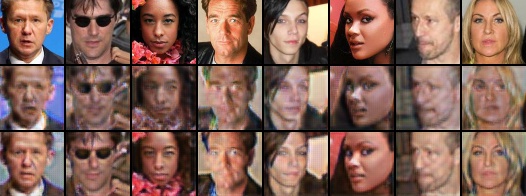}
        \put(-14.5,30){\footnotesize Truth}
        \put(-17,17){\footnotesize Different}
        \put(-14,5){\footnotesize Same}
    \end{overpic}
}
\caption{Additional results of auxiliary dataset distribution shift on FORA performed on CIFAR-10 and CelebA.}
\label{fig:additional_iid}
\end{figure}
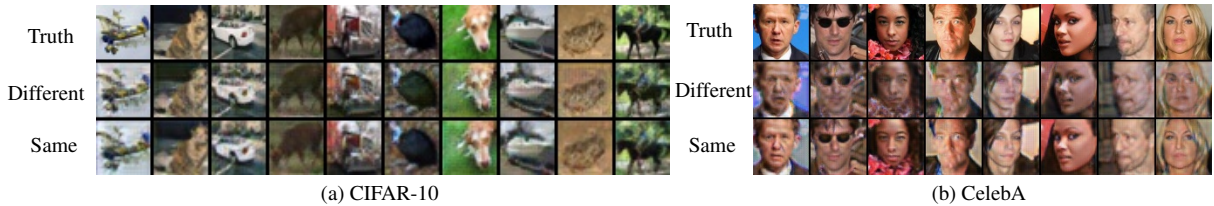

Then we present more detailed results of the impact of the auxiliary dataset on FORA.
\cref{fig:additional_class_missing} presents complete results for the absence of categories on FORA. 
\cref{fig:additional_quantity} displays additional results regarding the impact of the size of auxiliary datasets on FORA. 
\cref{fig:additional_iid} shows the overall results for the impact of auxiliary datasets distributions on FORA.

%--------------
%------------

%--------------
%------------

%--------------
%------------

\subsection{Effect of Substitute Client Structure}
\label{appendix_result_arc}

\begin{figure}[ht]
\centering
% \hspace{1.5pt}
\subfloat[CIFAR-10]{
\begin{overpic}[width=3.0in]{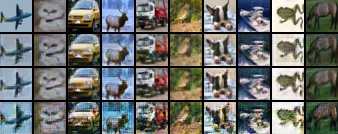}
    \put(-10,34){\footnotesize Truth}
    \put(-10,24){\footnotesize VGG}
    \put(-8,14){\footnotesize Res}
    \put(-11,4){\footnotesize Dense}
\end{overpic}
}
\hspace{20pt}
\subfloat[CelebA]{
\begin{overpic}[width=2.4in]{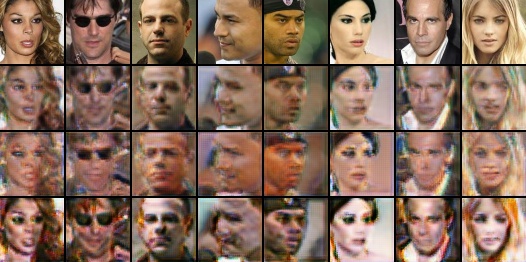}
    \put(-11,42){\footnotesize Truth}
    \put(-11,30){\footnotesize VGG}
    \put(-9,17){\footnotesize Res}
    \put(-12,5){\footnotesize Dense}
\end{overpic}
}
\caption{Additional results for FORA with varying substitute model architectures on CIFAR-10 and CelebA.}
\label{fig:additional_structure}
\end{figure}
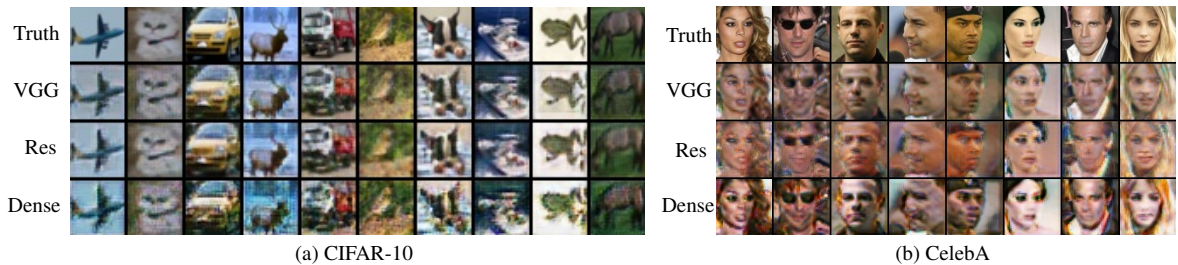

In addition to the quantified results in the main text, we also present the reconstructed images by FORA under different substitute client model architectures in  \cref{fig:additional_structure}.

%--------------
%--------------

\section{Defense Techniques}
\label{appendix_defense}

In this section, we first provide a detailed introduction to the several defense mechanisms evaluated in our experiments. Then we present the additional defense results of CelebA and CIFAT-10. We finally discuss two possible adaptive defenses against our proposed method FORA.

\subsection{Defense Details}
\label{appendix_defense_details}

\textbf{Gradients Scrutinizer.} 
Gradients Scrutinizer (GS) \cite{hust_ndss} is a defense method against the malicious attacker FSHA \cite{pasquini2021unleashing}. In normal SL, gradients returned by servers exhibit greater similarity within the same label compared to gradients from different labels.
However, in FSHA the client is trained as the encoder of an autoencoder to reconstruct training data without using target labels.
 As a result, gradients received by the client will not show notable distinctions between the same and different classes in FSHA.
 Based on this intrinsic difference, GS first computes the cosine similarity of gradients with the same label and those of different labels in each batch according to the received gradients.
 Subsequently, GS will calculate decision scores from three aspects: set gap, fitting error, and overlapping ratio, to distinguish hijacking servers from honest servers.
 If the detection score is above a set threshold, it is considered a normal SL. If it falls below the threshold, it is identified as a hijacking attack, and the training is stopped immediately.

The detection mechanism of GS depends on the model's classification ability. Therefore, in the early stages of GS, some batches should be skipped to avoid the detection being misguided by the model that has not been well-fitted.
Following this mechanism, we start GS at the 450th iteration in our experimental setup.

\textbf{Distance Correlation Minimization.} Distance Correlation \cite{vepakomma2020nopeek,szekely2007measuring,vepakomma2019reducing} is a defense method widely used in SL to measure and mitigate the correlation between smashed data and the raw input, thereby preventing server adversaries from reconstructing the original input data. The loss function for this approach is as follows:

\begin{equation}
\mathcal{L}={\alpha}{\cdot}{DCOR(x, F_c(x))}+(1-\alpha)\cdot{TASK(y, F_s(F_c(x)))}
\label{equ:dcor}
\end{equation}
where $DCOR$ represents the distance correlation metric, and $TASK$ denotes the classification loss between the true label $y$ and the model's prediction. By jointly minimizing the above loss, a better tradeoff can be achieved between preserving input data privacy and maintaining the utility of the model.

\textbf{Differential Privacy.} Differential Privacy was initially introduced to provide privacy guarantees for algorithms on aggregate databases \cite{dwork2006differential,dwork2014algorithmic}, and it was later applied to deep learning through DP-SGD \cite{abadi2016deep}. Differential privacy has found widespread usage in various scenarios \cite{lecuyer2019certified,shokri2015privacy}, not exclusively in SL. Following the approach described in PCAT \cite{gaopcat}, we employ DP-SGD in the client model. Specifically, the client receives gradients from the server, clips each gradient using a threshold value $C$, and adds random noise to it. The client then utilizes these protected gradients to update its model, thereby safeguarding the privacy of the subsequent smashed data transmitted to the server. Different combinations of the clipping threshold $C$ and noise scale $\sigma$ yield varying privacy budgets $\epsilon$ and levels of accuracy reduction.

\textbf{Noise Obfuscation.} Titcombe et al. \cite{titcombe2021practical} proposed an approach where additive Laplacian noise was directly added to smashed data before transmitting it to the server to defend against input reconstruction. This randomness introduces a higher level of complexity for adversaries, making it more challenging for them to infer the mapping between the smashed data and the private input.

\subsection{More Defense Results and Possible Adaptive Defenses}
\label{appendix_defense_celeba}
\textbf{Attack Results of CelebA.} The limited effectiveness of these defenses on CelebA is illustrated in \cref{tab:defense_celeba}.
In comparison to CIFAR-10, CelebA shows a more robust performance in terms of test accuracy. This is because CelebA is employed for a simpler binary classification task (smile classification), making the model more easily convergent even in the presence of defense methods.

\begin{table}[h]
\centering
\caption{Results on CIFAR-10 at layer 2 with smaller $\epsilon$.}
\resizebox{0.5\linewidth}{!}{
\begin{tabular}{c c cccc@{\hspace{0pt}} c@{\hspace{0pt}} c@{\hspace{0pt}} c@{\hspace{0pt}}}
    \hline
    DP ($\epsilon$) & Test Acc (\%) & SSIM$\uparrow$ & PSNR$\uparrow$ & LPIPS$\downarrow$
    & \begin{minipage}[c]{0.10\columnwidth}
    \centering
    {\includegraphics[width=1.0\textwidth]{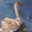}}
    \end{minipage}
    & \begin{minipage}[c]{0.10\columnwidth}
    \centering
    {\includegraphics[width=1.0\textwidth]{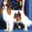}}
    \end{minipage}
    & \begin{minipage}[c]{0.10\columnwidth}
    \centering
    {\includegraphics[width=1.0\textwidth]{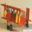}}
    \end{minipage}
    & \begin{minipage}[c]{0.10\columnwidth}
    \centering
    {\includegraphics[width=1.0\textwidth]{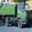}}
    \end{minipage}
    \\
    % \hline
    % \multicolumn{5}{c}{\textbf{DP ($\epsilon$)}} & & & &
    % \\
    \hline
    6 & 61.61\% & 0.590 & 17.49 & 0.496
    & \begin{minipage}[c]{0.10\columnwidth}
    \centering
    {\includegraphics[width=1.0\textwidth]{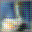}}
    \end{minipage}
    & \begin{minipage}[c]{0.10\columnwidth}
    \centering
    {\includegraphics[width=1.0\textwidth]{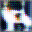}}
    \end{minipage}
    & \begin{minipage}[c]{0.10\columnwidth}
    \centering
    {\includegraphics[width=1.0\textwidth]{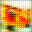}}
    \end{minipage}
    & \begin{minipage}[c]{0.10\columnwidth}
    \centering
    {\includegraphics[width=1.0\textwidth]{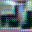}}
    \end{minipage}
    \\
    1 & 61.17\% & 0.582 & 17.43 & 0.522
    & \begin{minipage}[c]{0.10\columnwidth}
    \centering
    {\includegraphics[width=1.0\textwidth]{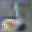}}
    \end{minipage}
    & \begin{minipage}[c]{0.10\columnwidth}
    \centering
    {\includegraphics[width=1.0\textwidth]{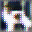}}
    \end{minipage}
    & \begin{minipage}[c]{0.10\columnwidth}
    \centering
    {\includegraphics[width=1.0\textwidth]{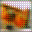}}
    \end{minipage}
    & \begin{minipage}[c]{0.10\columnwidth}
    \centering
    {\includegraphics[width=1.0\textwidth]{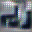}}
    \end{minipage}
    \\
    \hline
\end{tabular}
}
\label{tab:dp}
\end{table}

\begin{table}[b]
\centering
\caption{Effect of utility and FORA performance against distance correlation minimization, differential privacy and noise obfuscation on CelebA at layer 2.}
\resizebox{0.55\linewidth}{!}{
\begin{tabular}{c c ccc c@{\hspace{0pt}} c@{\hspace{0pt}} c@{\hspace{0pt}} c@{\hspace{0pt}}}
    \hline
    \makecell{Defense \\ Hyperparam} & Test Acc (\%) & SSIM$\uparrow$ & PSNR$\uparrow$ & LPIPS$\downarrow$
    & \begin{minipage}[c]{0.09\columnwidth}
    \centering
    {\includegraphics[width=1.0\textwidth]{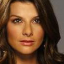}}
    \end{minipage}
    & \begin{minipage}[c]{0.09\columnwidth}
    \centering
    {\includegraphics[width=1.0\textwidth]{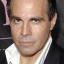}}
    \end{minipage}
    & \begin{minipage}[c]{0.09\columnwidth}
    \centering
    {\includegraphics[width=1.0\textwidth]{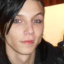}}
    \end{minipage}
    & \begin{minipage}[c]{0.09\columnwidth}
    \centering
    {\includegraphics[width=1.0\textwidth]{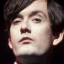}}
    \end{minipage}
    \\
    \hline
    \makecell{0 \\(w/o defense)} & 91.91 & 0.476 & 17.11 & 0.381
    & \begin{minipage}[c]{0.09\columnwidth}
    \centering
    {\includegraphics[width=1.0\textwidth]{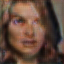}}
    \end{minipage}
    & \begin{minipage}[c]{0.09\columnwidth}
    \centering
    {\includegraphics[width=1.0\textwidth]{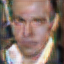}}
    \end{minipage}
    & \begin{minipage}[c]{0.09\columnwidth}
    \centering
    {\includegraphics[width=1.0\textwidth]{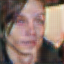}}
    \end{minipage}
    & \begin{minipage}[c]{0.09\columnwidth}
    \centering
    {\includegraphics[width=1.0\textwidth]{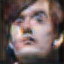}}
    \end{minipage}
    \\
    \hline
    \multicolumn{5}{c}{\textbf{DCOR ($\alpha$)}} & & & &
    \\
    \hline
    0.2 & 91.96 & 0.407 & 15.46 & 0.470
    & \begin{minipage}[c]{0.09\columnwidth}
    \centering
    {\includegraphics[width=1.0\textwidth]{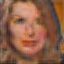}}
    \end{minipage}
    & \begin{minipage}[c]{0.09\columnwidth}
    \centering
    {\includegraphics[width=1.0\textwidth]{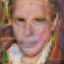}}
    \end{minipage}
    & \begin{minipage}[c]{0.09\columnwidth}
    \centering
    {\includegraphics[width=1.0\textwidth]{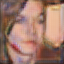}}
    \end{minipage}
    & \begin{minipage}[c]{0.09\columnwidth}
    \centering
    {\includegraphics[width=1.0\textwidth]{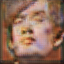}}
    \end{minipage}
    \\
    0.5 & 92.24 & 0.355 & 14.58 & 0.428
    & \begin{minipage}[c]{0.09\columnwidth}
    \centering
    {\includegraphics[width=1.0\textwidth]{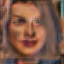}}
    \end{minipage}
    & \begin{minipage}[c]{0.09\columnwidth}
    \centering
    {\includegraphics[width=1.0\textwidth]{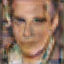}}
    \end{minipage}
    & \begin{minipage}[c]{0.09\columnwidth}
    \centering
    {\includegraphics[width=1.0\textwidth]{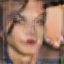}}
    \end{minipage}
    & \begin{minipage}[c]{0.09\columnwidth}
    \centering
    {\includegraphics[width=1.0\textwidth]{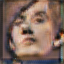}}
    \end{minipage}
    \\
    0.8 & 91.93 & 0.266 & 12.03 & 0.635
    & \begin{minipage}[c]{0.09\columnwidth}
    \centering
    {\includegraphics[width=1.0\textwidth]{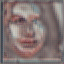}}
    \end{minipage}
    & \begin{minipage}[c]{0.09\columnwidth}
    \centering
    {\includegraphics[width=1.0\textwidth]{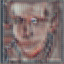}}
    \end{minipage}
    & \begin{minipage}[c]{0.09\columnwidth}
    \centering
    {\includegraphics[width=1.0\textwidth]{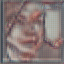}}
    \end{minipage}
    & \begin{minipage}[c]{0.09\columnwidth}
    \centering
    {\includegraphics[width=1.0\textwidth]{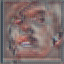}}
    \end{minipage}
    \\
    \hline
    \multicolumn{5}{c}{\textbf{DP ($\epsilon$)}}& & & &
    \\
    \hline
    $+\infty$ & 91.91 & 0.460 & 16.21 & 0.424
    & \begin{minipage}[c]{0.09\columnwidth}
    \centering
    {\includegraphics[width=1.0\textwidth]{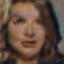}}
    \end{minipage}
    & \begin{minipage}[c]{0.09\columnwidth}
    \centering
    {\includegraphics[width=1.0\textwidth]{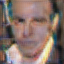}}
    \end{minipage}
    & \begin{minipage}[c]{0.09\columnwidth}
    \centering
    {\includegraphics[width=1.0\textwidth]{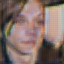}}
    \end{minipage}
    & \begin{minipage}[c]{0.09\columnwidth}
    \centering
    {\includegraphics[width=1.0\textwidth]{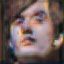}}
    \end{minipage}
    \\
    100 & 91.00 & 0.460 & 16.65 & 0.507
    & \begin{minipage}[c]{0.09\columnwidth}
    \centering
    {\includegraphics[width=1.0\textwidth]{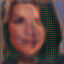}}
    \end{minipage}
    & \begin{minipage}[c]{0.09\columnwidth}
    \centering
    {\includegraphics[width=1.0\textwidth]{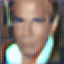}}
    \end{minipage}
    & \begin{minipage}[c]{0.09\columnwidth}
    \centering
    {\includegraphics[width=1.0\textwidth]{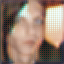}}
    \end{minipage}
    & \begin{minipage}[c]{0.09\columnwidth}
    \centering
    {\includegraphics[width=1.0\textwidth]{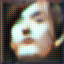}}
    \end{minipage}
    \\
    10 & 91.41 & 0.435 & 16.54 & 0.455
    & \begin{minipage}[c]{0.09\columnwidth}
    \centering
    {\includegraphics[width=1.0\textwidth]{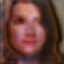}}
    \end{minipage}
    & \begin{minipage}[c]{0.09\columnwidth}
    \centering
    {\includegraphics[width=1.0\textwidth]{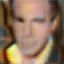}}
    \end{minipage}
    & \begin{minipage}[c]{0.09\columnwidth}
    \centering
    {\includegraphics[width=1.0\textwidth]{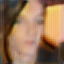}}
    \end{minipage}
    & \begin{minipage}[c]{0.09\columnwidth}
    \centering
    {\includegraphics[width=1.0\textwidth]{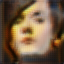}}
    \end{minipage}
    \\
    \hline
    \multicolumn{5}{c}{\textbf{NO ($\sigma$)}} & & & &
    \\
    \hline
    0.5 & 92.19 & 0.378 & 16.63 & 0.568
    & \begin{minipage}[c]{0.09\columnwidth}
    \centering
    {\includegraphics[width=1.0\textwidth]{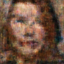}}
    \end{minipage}
    & \begin{minipage}[c]{0.09\columnwidth}
    \centering
    {\includegraphics[width=1.0\textwidth]{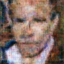}}
    \end{minipage}
    & \begin{minipage}[c]{0.09\columnwidth}
    \centering
    {\includegraphics[width=1.0\textwidth]{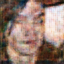}}
    \end{minipage}
    & \begin{minipage}[c]{0.09\columnwidth}
    \centering
    {\includegraphics[width=1.0\textwidth]{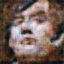}}
    \end{minipage}
    \\
    1.0 & 92.35 & 0.311 & 15.31 & 0.666
    & \begin{minipage}[c]{0.09\columnwidth}
    \centering
    {\includegraphics[width=1.0\textwidth]{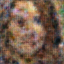}}
    \end{minipage}
    & \begin{minipage}[c]{0.09\columnwidth}
    \centering
    {\includegraphics[width=1.0\textwidth]{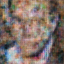}}
    \end{minipage}
    & \begin{minipage}[c]{0.09\columnwidth}
    \centering
    {\includegraphics[width=1.0\textwidth]{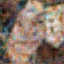}}
    \end{minipage}
    & \begin{minipage}[c]{0.09\columnwidth}
    \centering
    {\includegraphics[width=1.0\textwidth]{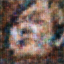}}
    \end{minipage}
    \\
    2.0 & 92.40 & 0.140 & 12.61 & 0.704
    & \begin{minipage}[c]{0.09\columnwidth}
    \centering
    {\includegraphics[width=1.0\textwidth]{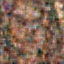}}
    \end{minipage}
    & \begin{minipage}[c]{0.09\columnwidth}
    \centering
    {\includegraphics[width=1.0\textwidth]{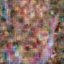}}
    \end{minipage}
    & \begin{minipage}[c]{0.09\columnwidth}
    \centering
    {\includegraphics[width=1.0\textwidth]{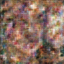}}
    \end{minipage}
    & \begin{minipage}[c]{0.09\columnwidth}
    \centering
    {\includegraphics[width=1.0\textwidth]{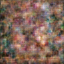}}
    \end{minipage}
    \\
    \hline
\end{tabular}
}
\label{tab:defense_celeba}
\end{table}
%--------------------------------------------------------------

\textbf{Results with Smaller $\epsilon$ on CIFAR-10.}
Table \ref{tab:dp} presents the results with smaller $\epsilon$ on CIFAR-10. We observe an interesting phenomenon: as the applied noise increased, there is a nonlinear relationship with the defense results. The possible reason is that the noise can only act on partial gradients (client), limiting its effectiveness.

\textbf{Possible Adaptive Defenses.}
We discuss two potential adaptive defenses.
One is that the client adopts an adversarial loss to enhance robustness against DRA \cite{adversarial_learning,li2022ressfl}. 
Though adversarial learning proves effective against certain known attacks, client should carefully consider the additional training overhead and  utility degradation it introduces.
Another promising approach is to craft noise against FORA to increase the inconsistency between client and substitute client in feature space, which would make attack more difficult \cite{wang2023privacy}.

\section{Limitation and Future Work}
\label{more_discussions}
In this section, we discuss the limitations of our proposed method FORA and some possible enhancements for future work. Previous work and FORA lack sufficient experiments on larger models and datasets \eg vision transformer \cite{vision}, so we encourage future work to pay more attention on larger models. 
Additionally, although FORA only requires auxiliary data of the same type to launch an attack, exploring how to reconstruct client inputs in more challenging scenarios, such as when attackers do not know the data type, is also unsolved.
We hope our work can contribute to better exploring  vulnerabilities of SL and raising awareness of privacy issues within the community.

\section{Label-Protected SL}
\begin{figure}[h]
\centering
\includegraphics[width=2.7in]{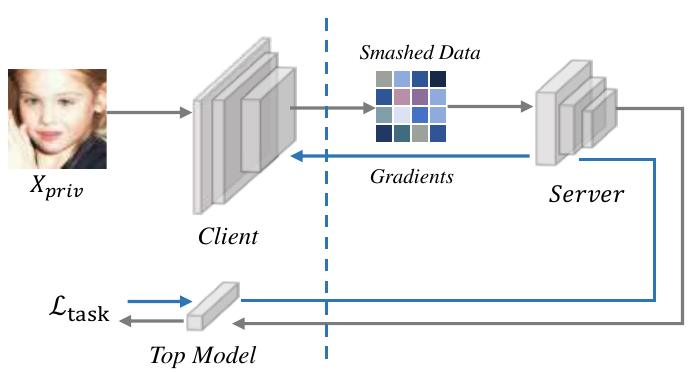}
\caption{Architecture of label-protected split learning.}
\label{fig:sl_label_protected}
\end{figure}

In label-protected SL \cite{vepakomma2018split} shown in \cref{fig:sl_label_protected}, besides the client model and server model, there is also a portion of model called the top model retained on the client. In this scenario, labels are treated as private information and kept locally on the client. 
Differing from label-share SL, the server model's results are forwarded to the top model for further forward propagation. The top model then calculates the loss using labels and received results, transferring relevant gradients to the server for parameter updates.

\end{document}